 \documentclass[onecolumn]{aa} 
\usepackage{graphicx}
\usepackage{txfonts}
\usepackage{bm}
\usepackage{longtable}
\usepackage{supertabular}
\begin{document}

\title{Equation of state of dense nuclear matter and neutron star structure 
       from nuclear chiral interactions}

 \author{Ignazio Bombaci\inst{1, 2}\fnmsep\thanks{email: ignazio.bombaci@unipi.it}
          \and
         Domenico Logoteta\inst{2}\fnmsep\thanks{email: domenico.logoteta@infn.pi.it}
           }
 \institute {Dipartimento di Fisica ``E. Fermi'', Universit\`a di Pisa, 
             Largo B. Pontecorvo, 3 I-56127 Pisa, Italy\\
             \and
            INFN, Sezione di Pisa, Largo B. Pontecorvo, 3 I-56127 Pisa, Italy\\
          }
\date{Received .........;  accepted ......}

 \abstract 
  {}
{We report a new microscopic equation of state (EOS) of dense symmetric nuclear matter, pure  
neutron matter, and asymmetric and $\beta$-stable nuclear matter at zero temperature using 
recent realistic two-body and three-body nuclear interactions derived in the framework of 
chiral perturbation theory (ChPT) and including the $\Delta(1232)$ isobar intermediate state. 
This EOS is provided  in tabular form and in parametrized form ready for use in numerical 
general relativity simulations of binary neutron star merging.    
Here we use our new EOS for $\beta$-stable nuclear matter to compute various structural properties 
of non-rotating neutron stars.} 
{The EOS is derived using the Brueckner--Bethe--Goldstone quantum many-body theory in the 
Brueckner--Hartree--Fock approximation. Neutron star properties are next computed 
solving numerically the Tolman--Oppenheimer--Volkov structure equations. }
{Our EOS models are able to reproduce the empirical saturation point of symmetric nuclear matter, 
the symmetry energy $E_{sym}$, and its slope parameter $L$ at the empirical saturation density $n_{0}$.     
In addition, our EOS models are compatible with experimental data from collisions between heavy nuclei 
at energies ranging from a few tens of MeV up to several hundreds of MeV per nucleon. 
These experiments provide a selective test for constraining the nuclear EOS up to $\sim 4 n_0$. 
Our EOS models are consistent with present measured neutron star masses and particularly with the 
mass $M = 2.01 \pm 0.04 \, M_{\odot}$ of the neutron stars in PSR~J0348+0432. }
{}
\authorrunning{Bombaci and Logoteta} 
\titlerunning{Equation of state of dense nucleon matter and neutron stars}

  \maketitle
%

\section{Introduction} 
With central densities exceeding  the density of atomic nuclei 
($2.6 \times 10^{14}{\rm g/cm}^3$) several times over, neutron stars (NSs) are the densest macroscopic objects in the universe. 
They thus represent incomparable natural laboratories that allow us to investigate the constituents of matter 
and their interactions under extreme conditions that cannot be reproduced in any terrestrial laboratory, 
and to explore the phase diagram of quantum chromodynamics (QCD) in a region that is presently 
inaccessible to numerical calculations of QCD on a space--time lattice \citep{delia2003,gupta2010,FH2011}.   

The global properties of NSs (mass, radius, maximum mass, maximum spin frequency, etc.)  
and their internal composition (constituent particle species and possible different phases of matter)  
primarily depend on the equation of state (EOS) of strong interacting matter \citep{prak97,Latt-Prak-2016} 
{i.e.,} on the thermodynamical relation between the matter pressure, energy density, and temperature.  
The EOS of dense matter is also a basic ingredient for modeling various astrophysical phenomena related 
to NSs, such as core-collapse supernovae (SNe) \citep{oertel2017} and binary neutron star (BNS) mergers. 

Determining the correct EOS model that  describes NSs is a fundamental problem of nuclear and 
particle physics and of astrophysics, and major efforts have been made during the last few decades 
to solve it by measuring different NS properties using the data collected by various generations 
of X-ray and $\gamma$-ray satellites and by ground-based radio telescopes.  

The recent detection of four gravitational wave events \citep{gw1,gw2,gw3,gw4} 
caused by binary black hole mergers, but in particular the very recent detection    
of gravitational wave signals from a binary neutron star merger \citep{gw5}, is giving a big boost to 
the research on dense matter physics. 
The gravitational wave signal, especially from the BNS post-merger phase, offers  a 
unique opportunity to test different dense matter EOS models 
\citep{shibata2005,BJ2012,Takami2014,Bernuzzi2015,Sekiguchi2016,RT2016,Bauswein2016,endrizzi2016,
maione2016,ciolfi2017,Radice2017,piro2017}.  
Thus, gravitational wave astronomy will open a new window to explore matter under extreme conditions. 

As mentioned before, due to their large central densities, various ``exotic'' constituents, 
for example hyperons \citep{hyp1,hyp2,hyp3,hyp4} or a quark deconfined phase of 
matter \citep{gle96,bombaci2008,log2012,bl2013,bombaci2016}, are expected in neutron star interiors.  

In the present work we consider the more traditional view where  the core of a NS 
is modeled as a uniform charge-neutral fluid of neutrons, protons, electrons, and muons in equilibrium 
with respect to the weak interaction ($\beta$-stable nuclear matter).      
Even in this ``simplified'' picture, the determination of the EOS from the underlying  
nuclear interactions remains a challenging theoretical problem.  
In fact, it is necessary to calculate  
the EOS to extreme  conditions of high density and high neutron-proton asymmetry, 
{i.e.,} in a regime where the EOS is poorly constrained by nuclear data and experiments.  
The nuclear symmetry energy is thus one of the most important quantities that controls the composition
and the pressure of $\beta$-stable nuclear matter \citep{bl91,zuo14}, and consequently many NS attributes  
such as the radius, moment of inertia, and crustal properties \citep{latt14,Latt-Prak-2016}.  

Another important issue is related to the role of three-nucleon forces (TNFs) on the EOS, 
particularly at high density. 
In fact, it is well known that TNFs are essential in order to reproduce the experimental binding energy of 
few-nucleon systems (A = 3, 4)  and the empirical saturation point   
($n_{0} = 0.16 \pm  0.01~{\rm fm}^{-3}$, $E/A|_{n_0} = -16.0 \pm 1.0~{\rm MeV}$) 
of symmetric nuclear matter.   
As shown by several microscopic calculations \citep{wff88,bbb97,apr98,li-schu_08} of the EOS of $\beta$-stable nuclear matter based on realistic nucleon-nucleon (NN) interactions supplemented with TNFs, it is possible 
to obtain NSs with maximum mass
{\footnote{Stellar masses are given in units of  solar mass $M_\odot = 1.989 \times 10^{33}$~g}\,}     
$M_{max} \sim 2~M_{\odot}$, thus in agreement with currently 
measured masses.  
However, the value of $M_{max}$ depends in a sensitive manner on the TNFs strength at high 
density \citep{li-schu_08}, thus indicating that the properties of  few-body nuclear systems  and of nuclear matter 
saturation  cannot be used to constrain TNFs at high density.  

Recent years have witnessed a significant progress in the description of nuclear interactions.  
In fact, the  chiral effective field theory (ChEFT) has opened a new avenue for the description 
of nuclear interactions \citep{weinb1,weinb2,weinb3,weinb4,epel09,machl11} and nuclear systems consistent with QCD, the fundamental theory of the strong interaction. 
The significant advantage of using this method consists in the fact that two-body, three-body, and even 
many-body nuclear interactions can be calculated perturbatively, i.e., order by order, according to 
a well-defined scheme based on a low-energy effective QCD Lagrangian that retains the symmetries of QCD 
and in particular the approximate chiral symmetry. 
Within this chiral perturbation theory (ChPT) the details of the QCD dynamics are contained in parameters known as 
the low-energy constants (LECs), 
which are fixed by low-energy experimental data. 

Recently \cite{maria_local} have formulated a fully local in coordinate-space two-nucleon  
chiral potential which includes the $\Delta(1232)$ isobar (hereafter the $\Delta$ isobar) intermediate state. 
This new potential represents the fully local version of the minimally non-local chiral interaction reported 
in \cite{maria15}. It has been pointed out by various authors \citep{kaiser98,krebs07} that 
a $\Delta$-full ChPT has an improved convergence with respect to the  $\Delta$-less ChPT.  
In addition, the $\Delta$-full ChPT naturally leads to TNFs induced by two-pion exchange 
with excitation of an intermediate $\Delta$, the celebrated Fujita--Miyazawa 
three-nucleon force \citep{fujita57}.

In this work, we present a new microscopic EOS of dense symmetric nuclear matter (SNM), 
pure neutron matter (PNM), and asymmetric and $\beta$-stable nuclear matter at zero temperature  
using the local chiral potential by \cite{maria_local} 
supplemented with TNFs \citep{logoteta16b} and employing the  
Brueckner--Bethe--Goldstone (BBG) \citep{bbg1,bbg2} many-body theory within the 
Brueckner--Hartree--Fock (BHF) approximation. 
This zero temperature EOS is provided both in tabular form and in parametrized form ready 
for use in numerical general  relativity simulations of binary neutron star merging  
after being supplemented by a thermal contribution as described  in 
\citep[e.g.,][]{shibata2005,BJ2012,Takami2014,Bernuzzi2015,RT2016,Bauswein2016,endrizzi2016,maione2016,ciolfi2017}.

In addition, we use our new EOS for $\beta$-stable nuclear matter to compute various structural properties 
of non-rotating neutron stars. 
The present work represents a development and an extension to high nuclear baryon densities ($n > 2.5 n_0$) 
relevant for astrophysical applications with respect to our previous works 
\citep{logoteta15,logoteta16a,logoteta16b} where ChPT nuclear interactions are used 
in BHF calculations of nuclear matter properties around the empirical saturation density.   

The paper is organized as follows: in  Section 2 we present the nuclear interactions we have considered; 
in Section 3 we describe the BBG many-body theory and  discuss the inclusion of TNFs in this framework; 
in Section 4 we present our results for the EOS of SNM and PNM;  in Section 5 we report the 
calculated symmetry energy and the EOS for asymmetric and $\beta$-stable nuclear matter; 
in Section 6 we present various neutron star properties, as calculated with our new EOS;  
in the last section we summarize our main results.

\section{Nuclear interactions in chiral perturbation theory}

In this section we briefly describe the specific interactions we have employed in the present work. 
Among the wide variety of nuclear interactions derived in the framework of ChPT,  
for the two-body nuclear interaction, we have used the fully local chiral potential at the  
next-to-next-to-next-to-leading-order (N3LO) of ChPT, including $\Delta$ isobar excitations in 
intermediate state (hereafter N3LO$\Delta$) recently proposed by \cite{maria_local}.  
This potential was originally  presented in \cite{maria15} in a minimal non-local form. 
We note that \cite{maria_local} report different parametrizations of the local potential obtained 
by fitting the low-energy NN experimental data using different long- and short-range cutoffs.   
In the calculations presented in this work, we use the {\it model~b} described in  \cite{maria_local} 
(see their Table II), which fits the Granada database \citep{granada} of proton-proton ($pp$) and 
neutron-proton ($np$) scattering data up to an energy of $125$ MeV in the laboratory reference frame 
and has a $\chi^2$/datum$\sim$ $1.07$. 

There is a great deal of experimental evidence 
that the $\Delta$ isobar plays an important role in nuclear  
processes. For instance the excitation of the $\Delta$ isobar is needed to reproduce the observed 
energy spectra of low-lying states in $s$- and $p$-shell nuclei and to reproduce the correct spin-orbit 
splitting of P-wave resonances in low-energy $n$-$\alpha$ scattering
\citep{pieper2001,nollett2007,carlson2015}.  
It is consequently very important to test this new chiral nuclear potential \citep{maria_local} also in 
nuclear matter calculations at high density for astrophysical applications. 

For the TNF, we have used the potential by \cite{epel_N2LO} calculated at the 
next-to-next-to-leading-order (N2LO) of ChPT  in its local version given by \cite{navra_N2LOL}. 
The N2LO TNF depends on the parameters $c_1$, $c_3$, $c_4$, $c_D$, and $c_E$, i.e., the LECs.   
The N2LO three-nucleon interaction keeps the same operatorial structure, 
including or not the $\Delta$ degrees of freedom \citep{krebs07}.   
We note that the constants $c_1$, $c_3$, and $c_4$ are already fixed at the two-body level by 
the N3LO interaction.    
However, when including the $\Delta$ isobar in the three-body potential, the parameters $c_3$ and $c_4$ 
take additional contribution from the Fujita--Miyazawa diagram.    
This diagram appears at the next-to-leading-order (NLO) of ChPT and it is clearly not present 
in the theory without the $\Delta$ (see discussion in \citealt{logoteta16b} for more details).  

The values of the LECs $c_i$  in the TNF used in the present work are reported in 
Table\ \ref{tab1} for two different parametrizations.  
As mentioned previously, the parameters $c_1$, $c_3$, and $c_4$ are already fixed by 
the NN and $\pi$N (pion--nucleon) interaction, while the parameters $c_D$ and $c_E$ are not determined 
at the two-body level and thus  have to be set by reproducing some specific observable 
of few-body nuclear systems or by reproducing the empirical saturation point of SNM.    

The first TNF parametrization (hereafter N2LO$\Delta1$) was  determined in \cite{logoteta16b}, 
fitting the LECs $c_D$ and $c_E$ to get a good saturation point for SNM.   
For the second TNF parametrization (hereafter N2LO$\Delta2$) the values of $c_D$ and $c_E$ were  
set in order to reproduce the $^3$H binding energy \citep{logoteta16b}.  


 \begin{table} 
\caption{Values of the low-energy constants (LECs) of the TNF models used in the present calculations.} 
\label{tab1}
\small
\centering
\begin{tabular}{l|cccccc}
\hline 
\hline
        & $c_D$ & $c_E$  & $c_1$ & $c_3$  & $c_4$   \\               
\hline
  N2LO$\Delta1$      & -0.10 &   1.30   & -0.057 & -3.63 & 3.14  \\ 
  N2LO$\Delta2$      & -4.06 &   0.37   & -0.057 & -3.63 & 3.14  \\ 
\hline
 \end{tabular}
\tablefoot{In the first and  second rows, we list the parametrizations of the N2LO three-body force 
with the $\Delta$ isobar excitations.  
The values $c_1$, $c_3$, and $c_4$ have been kept fixed. 
The LECs $c_1$, $c_3$, and $c_4$ are expressed in GeV$^{-1}$, whereas $c_D$ and $c_E$ are dimensionless.}
 \end{table} 

\section{The Brueckner--Bethe--Goldstone many-body theory}

The Brueckner--Bethe--Goldstone (BBG) many-body theory is based on a linked cluster expansion 
(the hole-line expansion) of the energy per nucleon $\widetilde E \equiv E/A$ of nuclear matter.   
The various terms of the expansion can be represented by Goldstone diagrams grouped according 
to the number of independent hole-lines ({i.e.,} lines representing empty single particle states 
in the Fermi sea). 
The basic ingredient in this approach is the Brueckner reaction matrix $G$, which sums, in a closed form, 
the infinite series of the so-called ladder-diagrams and takes into consideration  the short-range 
strongly repulsive part of the nucleon-nucleon interaction. 

In the general case of asymmetric nuclear matter with neutron number density $n_n$, 
proton number density $n_p$, total nucleon number density $n = n_n + n_p$, and 
isospin asymmetry (asymmetry parameter),   
\begin{equation}
             \beta = \frac{n_n - n_p}{n} \, ,
\label{asympar}
\end{equation} 
the reaction matrix depends on the isospin 3{rd} components $\tau$ and $\tau^\prime$ of 
the two interacting nucleons.   
Thus, there are  different G-matrices describing the $nn$, $pp$, and $np$ in medium effective interactions.   
They are obtained by solving the generalized Bethe--Goldstone equations 
\begin{equation}
 G_{\tau \tau^\prime}(\omega) =  v 
  + v  \sum_{k_a,k_b} 
\frac{\mid\vec{k_a},\vec{k_b}\rangle \, Q_{\tau \tau^\prime}\, \langle\vec{k_a},\vec{k_b}\mid}
     {\omega - \epsilon_{\tau}(k_a) - \epsilon_{\tau^\prime}(k_b) + i\eta } \, G_{\tau \tau^\prime}(\omega) \;,
\label{bg}
\end{equation}
where ${v}$ is the bare NN interaction (or a density dependent two-body effective interaction when 
three-nucleon forces are introduced; see next section) and 
the quantity $\omega$ is the so-called starting energy. 
In the present work we consider spin unpolarized nuclear matter, 
thus in equation (\ref{bg}) and in the following equations we drop the spin indices to simplify the formalism
{\footnote{Spin polarized nuclear matter within the BHF approach has been considered  
by, e.g., \cite{vb02} and  \cite{bomb+06}.}.   
The operator 
$\mid\vec{k_a},\vec{k_b}\rangle Q_{\tau \tau^\prime}\langle \vec{k_a},\vec{k_b}\mid$ 
(Pauli operator) projects on intermediate scattering states in which the momenta $\vec{k_a}$ and $\vec{k_b}$
of the two interacting nucleons are above their respective Fermi momenta $k_{F}^{[\tau]}$ and 
$k_{F}^{[\tau^\prime]}$ since states with momenta smaller that these values are occupied by the nucleons 
of the nuclear medium.  
Thus the Bethe--Goldstone equation describes the scattering of two nucleons in the presence of other 
nucleons, and the Brueckner $G$ matrix represents the effective interaction between two nucleons 
in the nuclear medium and properly takes into account the short-range correlations arising from the 
strongly repulsive core in the bare NN interaction.  

The single-particle energy $\epsilon_\tau(k)$ of a nucleon ($\tau = $n, p) with momentum $\vec{k}$ 
and mass $m_\tau$ is given by
\begin{equation}
       \epsilon_{\tau}(k) = \frac{\hbar^2k^2}{2m_{\tau}} + U_{\tau}(k) \ ,
\label{spe}
\end{equation}
where $U_{\tau}(k)$ is a single-particle potential that represents the mean field felt by a nucleon 
due to its interaction with the other nucleons of the medium. 
In the BHF 
approximation of the BBG theory, $U_{\tau}(k)$
is calculated through the real part of the  on-energy-shell $G$-matrix \citep{BBP63,HM72} 
and is given by
\begin{equation}
U_{\tau}(k) = \sum_{\tau^\prime} \sum_{k'\leq k_{F}^{[\tau^\prime]}} 
               \mbox{Re} \ \langle \vec{k},\vec{k^\prime} \mid 
               G_{\tau\tau^\prime}(\omega^*) \mid \vec{k},\vec{k^\prime} \rangle_a  \;,
\label{spp}
\end{equation}
where the sum runs over all neutron and proton occupied states, 
$\omega^* = \epsilon_{\tau}(k) + \epsilon_{\tau'}(k'),$ and  the matrix elements are properly 
antisymmetrized. 
We make use of the so-called continuous choice \citep{jeuk+67,gra87,baldo+90,baldo+91} for 
the single-particle potential $U_{\tau}(k)$ when solving the Bethe--Goldstone equation.    
As shown by \cite{song98} and \cite{baldo00}, the contribution of the three-hole-line diagrams 
to the energy per nucleon $E/A$ is minimized in this prescription and 
a faster convergence of the hole-line expansion for $E/A$ is achieved with respect to the  
gap choice for $U_{\tau}(k)$.  

In this scheme Eqs.\ (\ref{bg})--(\ref{spp}) have to be solved self-consistently 
using an iterative numerical procedure. 
Once a self-consistent solution is achieved, the energy per nucleon of asymmetric nuclear matter is  
\begin{equation}
\widetilde{E}(n,\beta) = \widetilde{E}^{kin}(n,\beta) + \tilde{V}(n,\beta) \, ,
\label{en1}
\end{equation}
where 
%
\begin{eqnarray}
\widetilde{E}^{kin}(n,\beta) &=&  
 \frac{1}{A} \sum_{\tau}\sum_{k \leq k_{F}^{[\tau]} } \frac{\hbar^2k^2}{2m_{\tau}} = \nonumber \\
                                &=&  \widetilde{E}^{kin}_{0}(n) \,
 \frac{1}{2} \bigg\{ \frac{m_p}{m} (1+\beta)^{5/3} + \frac{m_n}{m} (1-\beta)^{5/3} \bigg\} 
\label{ekin}
\end{eqnarray} 
is the total kinetic energy per nucleon. 
In the previous expression, $m_n$ and $m_p$ respectively denote the neutron and proton masses,    
$m=(m_n+m_p)/2$ the average nucleon mass, and 
\begin{equation}
\widetilde{E}^{kin}_{0}(n) \equiv \widetilde{E}(n,\beta=0) 
      = \frac{3}{5} \frac{\hbar^2}{4 \mu} \Big( \frac{3\pi^2}{2}\Big)^{2/3} n^{2/3}
\label{ekin0}
\end{equation} 
 the kinetic energy per nucleon of SNM, 
with $\mu = m_n m_p/(m_n + m_p)$ being the reduced nucleon mass.     
The second term in Eq. (\ref{en1}) gives the potential energy contribution to total energy per nucleon.  
In the BHF approximation it can be written as 
\begin{equation} 
   \widetilde{V}(n,\beta) = 
          \frac{1}{2} \frac{1}{A}\sum_{\tau}\sum_{k \leq k_{F}^{[\tau]} } U_{\tau}(k)  \ .
\label{epot}
\end{equation}

In this approach the two-body interaction ${v}$ is the only physical input for 
the numerical solution Bethe--Goldstone equation.

As  is well known, within the most advanced non-relativistic quantum many-body approaches  
it is not possible to reproduce the empirical saturation point of symmetric nuclear matter 
when using two-body nuclear interactions only. 
In fact, the saturation points obtained using different NN potentials lie within a narrow 
band called  the {Coester band} \citep{coester70,day81}, with either a too large 
saturation density or a too small binding energy ($B = -E/A$) compared to the empirical values.   
In particular, SNM turns out to be overbound with a too large saturation density when using modern 
high-precision NN potentials, fitting NN scattering data up to energy 
of $350$ MeV, with a $\chi^2$ per datum next to $1$ \citep{ZHLi06}.  
As in the case of few-nucleon systems \citep{kalantar12,hammer13,binder16} and also for the nuclear matter 
case, TNFs are considered to be the missing physical effect of the whole picture.  
The inclusion of TNF is thus required in order to reproduce a realistic saturation 
point \citep{FP81,bbb97,apr98,Li2008,taranto13,zuo14}. 
In addition, TNFs are crucial in the case of dense $\beta$-stable nuclear matter to obtain a stiff 
EOS \citep{bbb97,apr98,li-schu_08,chamel11} compatible with the measured masses  
$M = 1.97 \pm 0.04 \, M_\odot$  \citep{demo2010} and 
$M = 2.01 \pm 0.04 \, M_\odot$  \citep{anto2013} 
of the neutron stars in PSR~J1614-2230 and PSR~J0348+0432, respectively. 
 
Within the BHF approach TNFs cannot be used directly in their original form  because it would be necessary to solve three-body Faddeev equations in the nuclear medium 
(Bethe--Faddeev equations) \citep{bethe65,rajaraman-bethe67}, and 
currently this is a task still far from being  achieved.  
To circumvent this problem an effective density dependent two-body force $v_{NN}^{eff}(n)$   
is built starting from the original three-body force by averaging over one of the three nucleons
\citep{loiseau,grange89}.      

In the present work, following \cite{holt}, we derive a density dependent  
effective NN force averaging the chiral N2LO$\Delta1$ and N2LO$\Delta2$ TNFs in the nuclear medium,  
as described in more detail in \cite{logoteta16b}.   

The Bethe--Goldstone equation (\ref{bg}) is then solved adding this effective density dependent 
two-body force to the bare NN interaction (the N3LO$\Delta$ interaction in our case).
It is important to note that when the original N2LO$\Delta1$ and N2LO$\Delta2$ TNFs are reduced 
to an effective density dependent two-body interaction $v_{NN}^{eff}(n)$, the only terms that survive 
in PNM after the average are those proportional to the LECs $c_1$ and $c_3$ \citep{logoteta16b}. 
Thus, the calculations using the models N3LO$\Delta$+N2LO$\Delta1$ and N3LO$\Delta$+N2LO$\Delta2$    
give the same results in PNM because they are not affected by the values of the LECs  
$c_D$ and $c_E$, and they have the same values for the LECs $c_1$ and $c_3$ (see Table\ \ref{tab1}).

\section{Equation of state for symmetric nuclear matter and pure neutron matter}

\begin{figure}[t]
\centering
\includegraphics[width = 7.8cm]{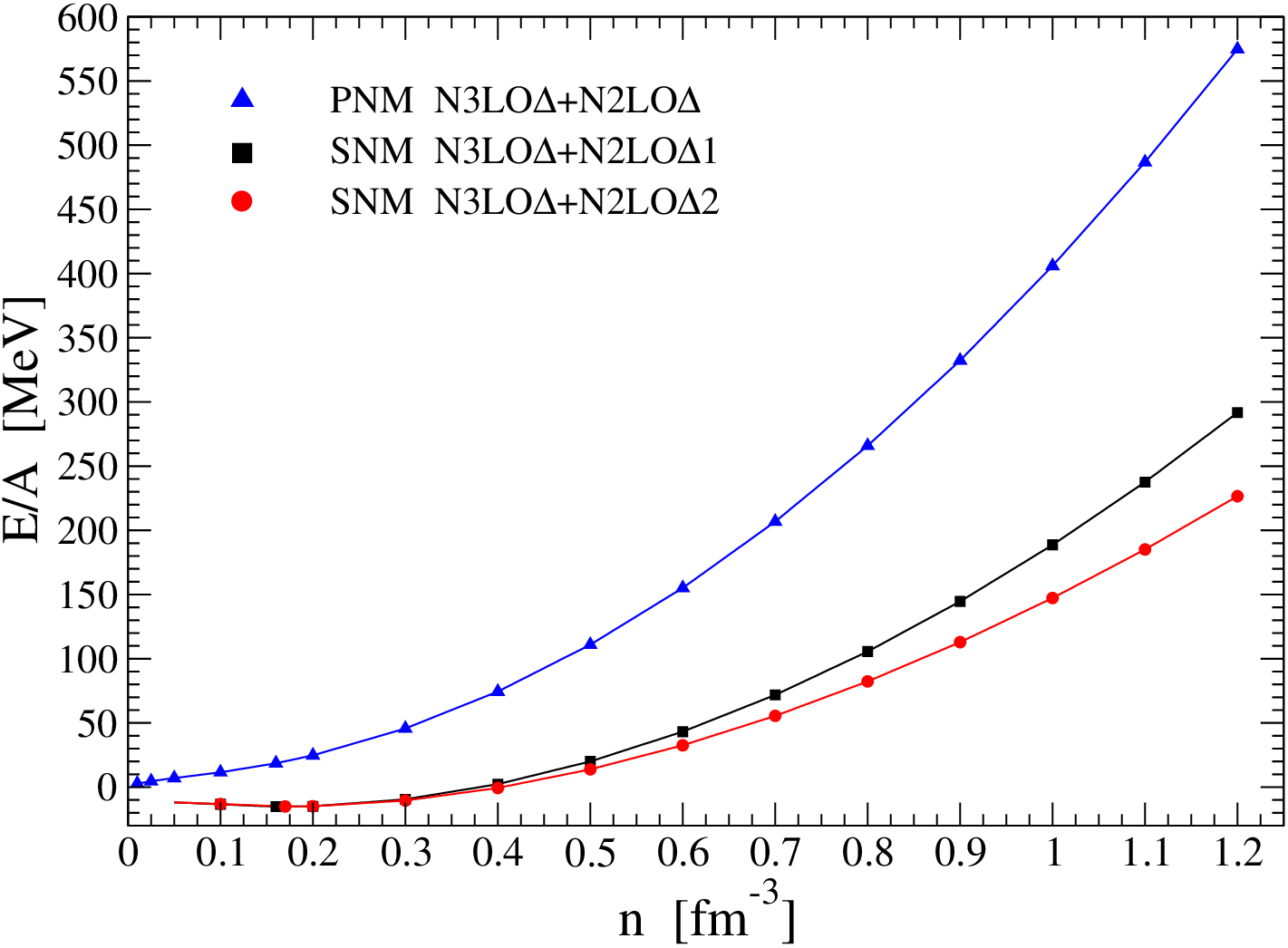} 
\caption{ Energy per particle of pure neutron matter (triangles) and symmetric nuclear matter 
(squares and circles) as a function of the nucleon number density for the interaction models considered 
in this work. The various symbols represent the results of our microscopic BHF calculations, 
whereas the lines represent the energy per particle obtained using the parametrization given by 
equations (\ref{SNM-fit}) and (\ref{PNM-fit}) for the potential energy contribution to $E/A$ 
for SNM and PNM.}
\label{fig1}
\end{figure}

In this section we present and discuss the results of our calculations for the equation of state,  
i.e., the energy per nucleon $E/A$ as a function of the baryon number density $n$, 
for SNM ($\beta = 0$) and PNM ($\beta = 1$)   
using the chiral nuclear interaction models and the BHF approach described in the previous pages.  
In this section of our work, we extend the nuclear matter calculations reported in \cite{logoteta16b} to high nucleon densities ($n > 0.4~\rm{fm^{-3}}$) 
relevant for neutron star physics, binary neutron star merging, and core-collapse supernovae. 
 
We note that we have to solve the general Eqs.\ (\ref{bg})--(\ref{spp}) even when we consider 
the case of SNM since the N3LO$\Delta$ two-nucleon interaction contains 
charge-independence breaking (CIB) 
and charge-symmetry breaking (CSB) terms (for a review, see, e.g., \citealt{miller2006})   
and since we consider the experimental values of the 
neutron and proton masses, i.e., we do not consider the approximation $m_n = m_p$.    

Making the usually adopted angular average of the Pauli operator and of the energy denominator  
\citep{gra87,baldo+91}, the Bethe--Goldstone equation (\ref{bg}) can be expanded in partial waves.  
In all the calculations performed in the present work, we have considered partial wave 
contributions up to a total two-body angular momentum $J_{max} = 8$.  
We have verified that the inclusion of partial waves with $J_{max} > 8$ does not appreciably 
change our results. For example, the relative change in the calculated BHF potential energy 
per nucleon (Eq.\,(\ref{epot})) in SNM at density $n = 1.0~{\rm fm}^{-3}$ when including 
partial wave contributions up to $J_{max} = 10$ is  
$(\widetilde{V}_{J_{max}=10} - \widetilde{V}_{J_{max}=8})/\widetilde{V}_{J_{max}=8} = 0.0035$.  
 
In Fig.\ \ref{fig1} we show the energy per nucleon of SNM obtained using the two parametrizations   
(see Table 1)  of the chiral N2LO$\Delta$ TNF, namely N2LO$\Delta1$ (squares) and 
N2LO$\Delta2$ (circles).   
The different symbols represent the results of our microscopic BHF calculations,   
whereas the lines represent the energy per particle obtained using the parametrization given by 
equations (\ref{SNM-fit}) and (\ref{PNM-fit}) for the potential energy contribution to $E/A$ for SNM and PNM 
discussed in the next section. 
It is apparent that at low density ($n < 0.3$~fm$^{-3}$) the two models produce almost identical results. 
At $n = 0.4$~fm$^{-3}$ the difference between the energy per nucleon originating from the two TNF models is 
$\sim 3$ MeV. This energy difference increases for increasing nucleon densities and is equal to 
$\sim 41.5$ MeV at $n = 1.0$~fm$^{-3}$.   
In the case of PNM the energy per particle (triangles in Fig.\ \ref{fig1}) for the two 
TNF models coincide because, as discussed in the previous section, neutron matter is not affected by terms proportional to the LECs $c_E$ and $c_D$. 

In Table\ \ref{tab2} we list the calculated values of the saturation points of SNM for the 
two interaction models considered in the present work. 
As we can see, the empirical saturation point of SNM,    
$n_{0} = 0.16 \pm  0.01~{\rm fm}^{-3}$, $E/A|_{n_0} = -16.0 \pm 1.0~{\rm MeV}$, 
is fairly well reproduced by our microscopic calculations. 
In Table\ \ref{tab2} we also report the nuclear symmetry energy, calculated as 
  $E_{sym}(n) = \widetilde{E}(n,\beta=1) - \widetilde{E}(n,\beta=0),$ 
and the symmetry energy slope parameter,     
\begin{equation}
       L = 3 n_{0} \frac{\partial E_{sym}(n)}{\partial n}\Big|_{n_{0}} 
\label{slope}
,\end{equation} 
at the calculated saturation density $n_0$ (second column in Table\ \ref{tab2}).  
Our calculated $E_{sym}(n_0)$ and $L$ are in a satisfactory agreement with the values obtained 
by other BHF calculations with two- and three-body interactions 
\citep{ZHLi06,li-schu_08,vidana09,vidana11} and with the values extracted from various 
nuclear experimental data \citep{latt14}.     

The incompressibility of SNM   
\begin{equation}
       K_\infty = 9 n_{0}^2 \, \frac{\partial^2 E/A}{\partial n^2}\Big|_{n_{0}}  
\label{incom}
\end{equation}
at the calculated saturation point for the interaction models used in the present work  
is given in the last column of Table\ \ref{tab2}.    
Our calculated values for $K_\infty$ underestimate the empirical value  
$K_\infty = 210 \pm 30$ MeV \citep{blaizot76} or more recently $K_\infty = 240 \pm 20$~MeV \citep{shlo06}   
extracted from experimental data of giant monopole resonance energies in medium-mass and heavy nuclei.
This is a common feature with many other BHF nuclear matter  
calculations with two- and three-body nuclear interactions \citep{li-schu_08,vidana09}. 

%
\begin{table*}  
\caption{Properties of nuclear matter for the interaction models used in this work.}    
\label{tab2}
\centering
\tiny
\begin{tabular}{cccccc}
\hline
\hline
Model & $n_0$(fm$^{-3}$) & $E/A$ (MeV) & $E_{sym}$ (MeV)  & $L$ (MeV)  & $K_\infty$ (MeV) \\
\hline
 N3LO$\Delta$+N2LO$\Delta1$ & 0.171  & -15.23   & 35.39   &  76.0   &  190 \\
 N3LO$\Delta$+N2LO$\Delta2$ & 0.176  & -15.09   & 36.00   &  79.8   &  176 \\
\hline
\end{tabular}
\tablefoot{Saturation density $n_0$ (Col. 2) and corresponding energy per nucleon $E/A$ (Col. 3) 
for symmetric nuclear matter; symmetry energy $E_{sym}$ (Col. 4); its slope parameter 
$L$ (Col. 5) and incompressibility $K_\infty$ (Col. 6) at the calculated saturation density.}

\end{table*}

\begin{figure}[t]
\centering
\vspace{1.1cm}
\includegraphics[width=8cm]{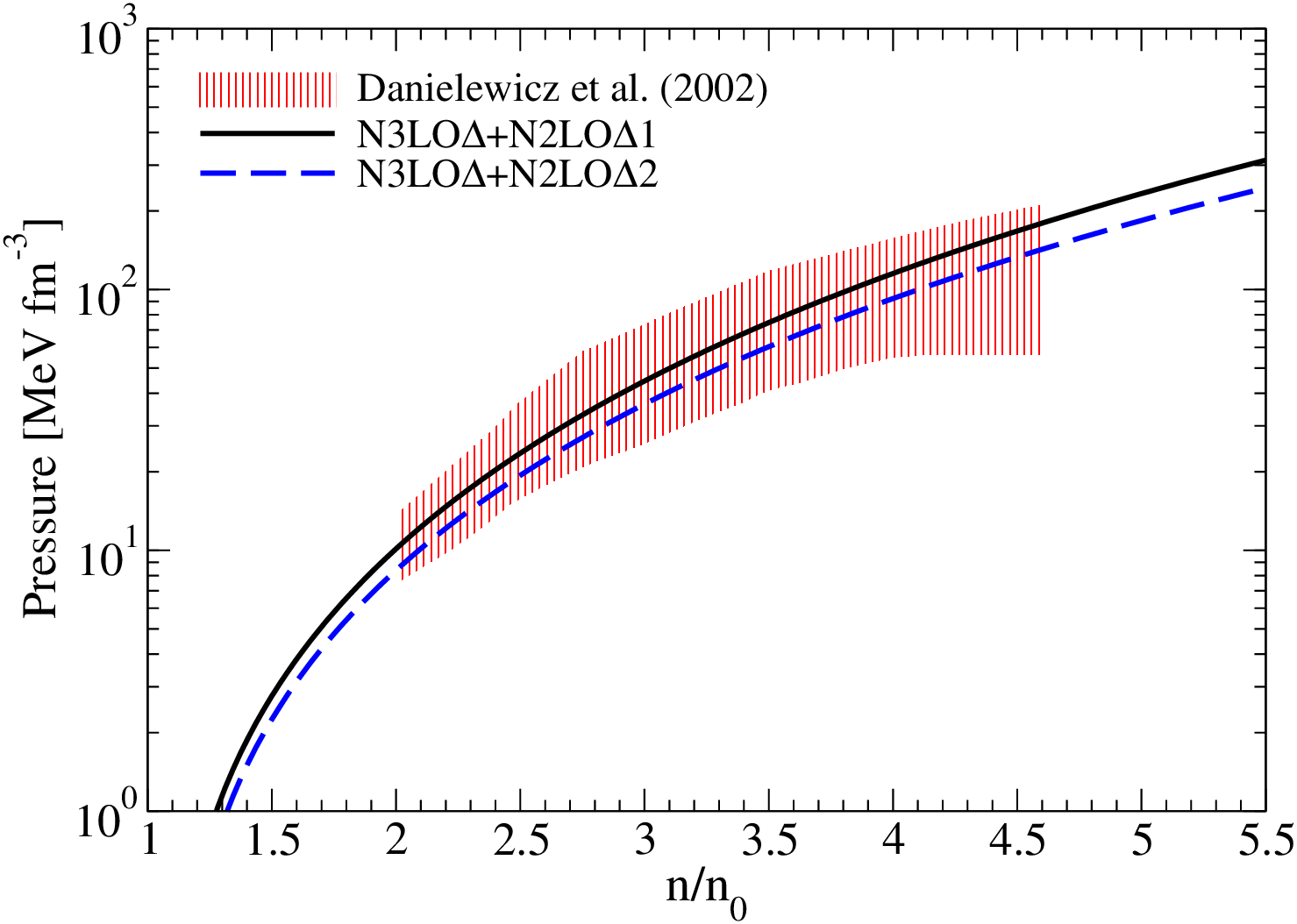} 
\caption{ Pressure of symmetric nuclear matter for the two interaction models used in this work. 
The red hatched area represents the region in the pressure--density plane for SNM  
which is consistent with the measured elliptic flow of matter in collision experiments 
between heavy atomic nuclei \citep{Dan2002}. 
}
\label{fig31}
\end{figure}

In addition to the empirical constraints at density around the saturation density $n_0$,  
the nuclear EOS can be tested using experimental data from collisions between heavy nuclei at energies ranging 
from a few tens of MeV up to several hundreds of MeV per nucleon. 
These collisions can compress nuclear matter up to $\sim 4 n_0$, thus giving valuable empirical information 
on the nuclear EOS at these supranuclear densities. 
Based on numerical simulations that reproduce the measured elliptic flow of matter in collision experiments 
between heavy nuclei, \cite{Dan2002} have been able to obtain a region in the pressure--density plane 
for SNM which is consistent with these elliptic flow experimental data. 
This region is represented by the red hatched area in Fig.\ \ref{fig31}.  
These collision experiments between heavy nuclei thus provide a selective test for constraining 
the nuclear EOS up to $\sim 4 n_0$. 
In the same figure, we show the pressure for our two EOS models for SNM obtained from the calculated 
energy per nucleon and using the standard thermodynamical relation 
\begin{equation}
         P(n)  = n^2 \frac{\partial (E/A)}{\partial n}\Big|_{A} \ . 
\label{PresSNM}
\end{equation}
As we can see our results are fully compatible with the empirical constraints given by \cite{Dan2002}. 

We want to emphasize that our BHF code, when used in conjunction with the N3LO$\Delta$ NN interaction plus 
our two parametrizations of the N2LO$\Delta$ TNF, reaches numerical convergence in the self-consistent 
scheme within a reasonable number of iterations (between $\sim 7$ and $14$) and up to the largest 
densities ($n \sim 1.2$~fm$^{-3}$) typical of neutron star maximum mass configurations.  
Thus, the nuclear matter EOS can be calculated fully microscopically up to these large densities.  
On the other hand, our BHF code does not reach convergence, already at density $\sim 0.5$ fm$^{-3}$, 
when used in conjunction with other interaction models derived at the same order of the $\Delta$-less 
ChPT \citep{logoteta16b}. Thus, in order to use these other interaction models for neutron star 
structure calculations, it is necessary to make a questionable extrapolation of 
the EOS to large densities.  

This important difference in the convergence of the BHF scheme with chiral interactions 
is related to  the inclusion of the $\Delta$ isobar both in the two- and three-nucleon potentials 
used in our present calculations.  
In fact, the $\Delta$-full ChPT has an improved convergence \citep{kaiser98,krebs07}
with respect to the  $\Delta$-less ChPT.

\section{Symmetry energy and EOS for asymmetric and 
                      $\beta$-stable nuclear matter}

The EOS of asymmetric nuclear matter can be calculated solving numerically 
Eqs.\ (\ref{bg})--(\ref{spp}) and (\ref{epot}) for various values of the asymmetry parameter 
($0 \leq \beta \leq 1$)  and for various densities ($ 0.5 \leq n/n_0 \leq 8$).    
These systematic calculations are not particularly demanding from a computational point of view;  
however, the use of these EOS tables is not ideal for applications to numerical simulations 
in general relativistic hydrodynamics.    
Thus, in the present work, in addition to providing EOS in tabular forms,  
we derive an EOS for asymmetric and $\beta$-stable nuclear matter at zero temperature 
in parametrized forms ready to be used in numerical simulations of   
binary neutron star merging.  

To this end, instead of using the general expression given in Eq. (\ref{epot}) 
for the BHF potential energy contribution $\widetilde{V}(n,\beta)$ to the energy per nucleon of 
asymmetric nuclear matter, we employ the so-called {parabolic approximation} in the asymmetry 
parameter $\beta$  \citep{bl91} 
\begin{equation}
\widetilde{V}(n,\beta) = \widetilde{V}_{0}(n) + E^{pot}_{sym}(n)\, \beta^2 \, ,
\label{en4}
\end{equation}
with $\widetilde{V}_{0}(n) \equiv \widetilde{V}_{0}(n, \beta=0)$ and $E^{pot}_{sym}(n)$ 
being the potential energy contribution to the energy per nucleon of SNM and to the 
total symmetry energy $E_{sym}$, respectively. 
Using Eq.\,(\ref{en4}), $E^{pot}_{sym}$ can be written as the difference between the potential energy 
contribution to the energy per nucleon of PNM and SNM, i.e.,
\begin{equation}
 E^{pot}_{sym}(n) = \widetilde{V}(n,\beta=1) - \widetilde{V}(n,\beta=0) \,.
\label{en5}
\end{equation}
It is important to note that the presence of tiny CSB and CIB terms in the nuclear interaction used in the 
present calculations could invalidate Eq. (\ref{en4}). 
For example, a CSB component in the NN interaction produces a linear (and more generally 
odd-power) $\beta$-term in Eq. (\ref{en4}) \citep{Haensel1997}.      
We have numerically checked the accuracy of Eq. (\ref{en4})      
for the N3LO$\Delta$ nucleon-nucleon interaction 
up to the high densities considered in the present calculations.   
Thus, in agreement with the results of \cite{Haensel1997} and \cite{muther1999}, 
the effects on the EOS of asymmetric nuclear matter and on the nuclear symmetry energy 
of CSB and CIB terms in the NN interaction are essentially negligible.  

The nucleon chemical potentials 
$\mu_{\tau}$ ($\tau =  n,\, p$), inclusive of the rest mass of the particle,  
can be thus written as 
\begin{equation}
\mu_{\tau}(n,\beta) = \frac{\partial \varepsilon_N}{\partial n_\tau} = 
                \mu_{\tau}^{kin}(n,\beta) + \mu_{\tau}^{pot}(n,\beta) + m_{\tau} c^2 \, ,
\label{chempot-2}
\end{equation} 
with 
\begin{equation}
  \mu_{\tau}^{kin}(n,\beta) = \frac{\hbar^2}{2m_\tau} \bigg(\frac{3\pi^2}{2}\bigg)^{2/3} \, n^{2/3} 
                 \Big(1 \pm \beta\Big)^{2/3} \, ,
\label{chempot-3}
\end{equation} 
\begin{equation} 
\mu_{\tau}^{pot}(n,\beta) = 
         \widetilde{V}_{0}(n) + n \frac{\partial \widetilde{V}_{0}}{\partial n} 
         \pm 2 E^{pot}_{sym}(n) \, \beta  + 
\bigg[ n \frac{\partial E^{pot}_{sym}(n)}{\partial n} - E^{pot}_{sym}(n) \bigg] \, \beta^2 \, ,
\label{chempot-4}
\end{equation} 
where the partial derivatives of the nucleonic energy density 
\begin{equation} 
\varepsilon_N(n,\beta) = n \widetilde{E}(n,\beta) + m_n n_n + m_p n_p
\label{endens1}
\end{equation} 
are taken at zero temperature and constant volume, and the upper and lower sign in equations 
(\ref{chempot-3}) and (\ref{chempot-4}) refers to neutrons and protons, respectively.  

 Subsequently, we parametrize the potential energy contribution to the energy per nucleon 
of SNM and PNM as 
\begin{equation}
\widetilde{V}_0(n) = \widetilde{V}(n,\beta=0) = a_0 n + b_0 n^{\gamma_0} + d_0 \,,
\label{SNM-fit}
\end{equation} 
\begin{equation}
\widetilde{V}_1(n) = \widetilde{V}(n,\beta=1) = a_1 n + b_1 n^{\gamma_1} + d_1  \,.
\label{PNM-fit}
\end{equation} 
We have fixed the values of the coefficients in Eqs. (\ref{SNM-fit}) and (\ref{PNM-fit}) 
fitting the results of our microscopic BHF calculations for SNM and PNM in the density range 
0.10--1.20~fm$^{-3}$.  
The coefficients given in Table\ \ref{tab3} fit the BHF results with a 
root mean square relative error ${\rm (RMSRE)} = 0.0069$ for both interactions in the case of SNM, 
and ${\rm RMSRE} = 0.0104$ in the case of PNM.  
The energy per particle $E/A$ corresponding to this parametrization is represented 
by the different lines in Fig.\ \ref{fig1}. 

Thus, using Eq.\, (\ref{en5}) the potential part of the symmetry energy 
can be written as
\begin{equation}
  E^{pot}_{sym}(n) = (a_1 - a_0) n +  b_1 n^{\gamma_1} - b_0 n^{\gamma_0} + d_1 - d_0 \,.
\label{en6}
\end{equation}
Equations (\ref{en4}), (\ref{SNM-fit}), and (\ref{en6}), together with Eqs. (\ref{ekin}) and (\ref{ekin0}) 
giving the kinetic energy per nucleon, provide our parametrized EOS for asymmetric nuclear matter. 

We next calculate the composition of $\beta$-stable nuclear matter, solving the equations 
for chemical equilibrium in neutrino-free matter 
($\mu_{\nu_e} = \mu_{\bar{\nu}_e} = \mu_{\nu_\mu} = \mu_{\bar{\nu}_\mu}$) 
(Prakash et al., 1997) at a given total nucleon density $n$  
\begin{equation}
\mu_n-\mu_p=\mu_e \;, \ \ \ \ \ \ \ \mu_e=\mu_{\mu}  
\label{beta1}
\end{equation}
and for charge neutrality 
\begin{equation}
   n_p = n_e +  n_{\mu} \,,
\label{beta2}
\end{equation}
with electrons and muons treated as relativistic ideal Fermi gases. 

\begin{table}
\caption{Coefficients of the parametrization  for the equation of state 
for symmetric nuclear matter (Eq. \ref{SNM-fit}) and  for pure matter (Eq. \ref{PNM-fit}). }
\label{tab3}
\small
\centering
\begin{tabular}{cccccc}
\hline
\hline
         Model (SNM)       &   $d_0$    &  $a_0$   &   $b_0$  & $\gamma_0$  \\
\hline
N3LO$\Delta$+N2LO$\Delta1$ & -9.22741   & -283.58  & 406.625  & 1.71844  \\
N3LO$\Delta$+N2LO$\Delta2$ & -8.62944   & -311.279 & 392.288  & 1.58626  \\
\hline
\\
\hline
\hline
 Model (PNM) & $d_1$  & $a_1$ & $b_1$  & $\gamma_1$  \\
\hline
N3LO$\Delta$+N2LO$\Delta1$ &  -0.877941  &  -208.176 & 496.125   &  1.81656  \\
N3LO$\Delta$+N2LO$\Delta2$ &  -0.877941  &  -208.176 & 496.125   &  1.81656  \\
\hline
\end{tabular}
\tablefoot{The coefficients $d_0$, $d_1$ are given in MeV; $a_0$, $a_1$ in MeV fm$^3$; 
and $b_0$,  $b_1$ in MeV fm$^{3 \gamma_0}$ and MeV fm$^{3 \gamma_1}$ respectively.}
\end{table}

%

The potential energy contribution to the nucleon chemical potential can thus be written as 
\begin{eqnarray}
\mu_{\tau}^{pot} &=& 2 a_0 n + (\gamma_0 + 1) b_0 n^{\gamma_0} + d_0 
                           \pm 2 E^{pot}_{sym}(n) \beta + \nonumber \\
                 &+&
[(\gamma_1 - 1) b_1 n^{\gamma_1} - (\gamma_0 - 1) b_0 n^{\gamma_0} - (d_1 - d_0)] \beta^2 \, ,
\label{chempot-5}
\end{eqnarray}
where the upper and lower sign refers to neutrons and protons, respectively.     
Consequently, the difference between the neutron and proton chemical potentials entering 
in the $\beta$-equilibrium  condition (\ref{beta1}) can be written as 
\begin{eqnarray}
\mu_n - \mu_p  &=& \frac{\hbar^2}{2m} \bigg(\frac{3\pi^2}{2}\bigg)^{2/3} \, n^{2/3} 
  \bigg\{ \frac{m}{m_n} \Big(1 + \beta\Big)^{2/3} - \frac{m}{m_p} \Big(1 - \beta\Big)^{2/3}\bigg\} + \nonumber \\
     &+& 4 E^{pot}_{sym}(n) \beta + (m_n - m_p) c^2 \, .
\label{mu_hat}
\end{eqnarray}
The composition of $\beta$-stable matter, {i.e.,} the particle fractions 
$x_i = n_i/n$ (with $i =$ n, p, $e^-$, $\mu^-$) calculated using the parametrization 
(\ref{SNM-fit}) and (\ref{PNM-fit}) of our microscopic calculations,  
is shown in Fig.\ \ref{part-fract}. 
The continuous (dashed) lines refer to the model N3LO$\Delta$+N2LO$\Delta1$ (N3LO$\Delta$+N2LO$\Delta2$).  
These results are in agreement with various other microscopic BHF calculations based on realistic nuclear   
interactions \citep{bbb97,burgio2011}.    

When the proton fraction $x_p = n_p/n$ is larger than a threshold value, denoted as $x_p^{durca}$, 
the direct URCA  processes 
$n \rightarrow p + e^- + \bar{\nu}_e\,$, 
$~ p + e^-  \rightarrow n +  \nu_e\,$
can occur in neutron star matter (Lattimer et al. 1991). 
 
In $\beta$-stable nuclear matter we can easily show that  
\begin{equation}
    x_p^{durca} = \frac{1}{1 + \big(1 + Y_e^{1/3}\big)^3} \, ,
\label{durca}
\end{equation}
where $Y_e = n_e/(n_e + n_\mu)$ is the leptonic electron fraction. 

The threshold proton fraction for direct URCA processes is depicted in Fig.\ \ref{part-fract} 
by the continuous line labeled  $x_p^{durca}$. 
Below the muon threshold density $x_p^{durca} = 1/9$.  

The calculated values for the threshold nucleon number density $n^{durca}$ for the 
occurrence of direct URCA  processes, for the corresponding proton fraction $x_p(n^{durca})$  
together with the corresponding neutron star gravitational mass $M^{durca} = M(n_c=n^{durca})$  
are reported in  Table\ \ref{durca2}.

\begin{table} 
\caption{Threshold values for the occurrence of direct URCA processes in dense $\beta$-stable nuclear matter.} 
\label{durca2}
\small
\centering
\begin{tabular}{cccc}
\hline
\hline
       Model  &   $n^{durca}$ (fm$^{-3}$) & $x_p(n^{durca})$ &   $M^{durca}/M_\odot$ \\
\hline
N3LO$\Delta$+N2LO$\Delta1$   & 0.361   &  0.1347  &  0.961    \\ 
N3LO$\Delta$+N2LO$\Delta2$   & 0.345   &  0.1343  &  0.862    \\ 
\hline
\end{tabular}
\tablefoot{Threshold baryon density $n^{durca}$ (given in fm$^{-3}$), proton fraction $x_p(n^{durca}),$ 
and neutron star gravitational mass $M^{durca} = M(n^{durca})$ for the considered EOS models.}
\end{table}

Once we have determined the particle fractions $x_i(n)$ in $\beta$-stable matter, 
the corresponding nucleonic contribution $\varepsilon_N(n)$ to the total energy density 
$\varepsilon(n) = \varepsilon_N(n) + \varepsilon_L(n)$
can be obtained using Eq.\,(\ref{endens1}).  
In addition, the nucleonic contribution $P_N(n)$ to the total pressure 
$P(n) = P_N(n) + P_L(n)$   
can be calculated using the thermodynamic relation 
\begin{equation}
    P_N = \mu_n n_n + \mu_p n_p - \varepsilon_N \, .
\label{EOS-beta}
\end{equation} 
Finally, the leptonic contributions $\varepsilon_L$ and $P_L$ to the total energy density and 
total pressure, respectively, are computed using the expressions for relativistic ideal Fermi gases  
with $m_e = 0.511$~MeV/c$^2$ and $m_\mu = 105.658$~MeV/c$^2$.

\begin{figure}[t]
\centering
\vspace{1.2cm}
\includegraphics[width=8cm]{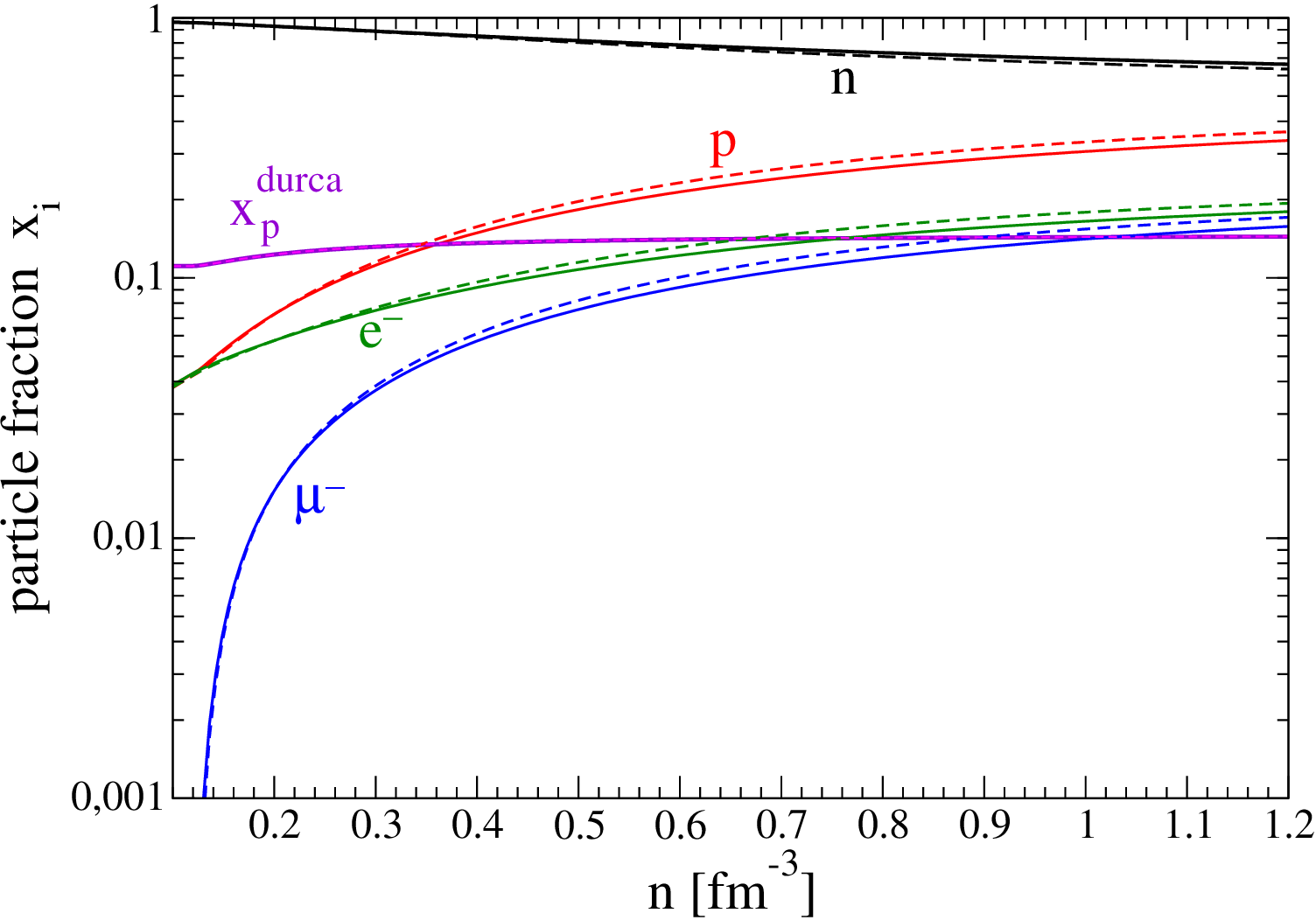}
\caption{Particle fractions of $\beta$-stable nuclear matter for the models 
described in the text.}
\label{part-fract}
\end{figure}

The resulting EOS for $\beta$-stable matter for the two considered nuclear interaction models 
is shown in Fig.\ \ref{eos_beta}. 
These results are consistent with those reported in Fig.\ \ref{fig1}. 
 
The EOS for $\beta$-stable matter is also reported in tabular form in the Appendix,  
where in addition to the baryon number density, energy density, and pressure, we  also list 
the proton fraction $x_p$ and the electron fraction $x_e$. The muon fraction is given by 
$x_\mu = x_p - x_e$. 

Our tabular EOS for $\beta$-stable matter can be reproduced in a simple and very accurate parametrized form. 
To this end we parametrize the total energy density $\varepsilon$  
(second column in Tables A.1 and A.2)  
as a function of the nucleon number density $n$ using the simple equation 
\begin{equation} 
    \varepsilon  = a n + b n^{\Gamma} \,.  
\label{beta-stableEOS_fit1}
\end{equation}
Then the total pressure can be deduced using the thermodynamical relation 
\begin{equation} 
       P = n \frac{\partial \varepsilon}{\partial n} - \varepsilon  
\label{pressure}
\end{equation}  
and is given by the polytrope 
\begin{equation} 
    P   =  (\Gamma - 1) b n^{\Gamma} =  K \rho_{rm}^{\Gamma}     
\label{beta-stableEOS_fit2}
,\end{equation} 
where $\rho_{rm} = (a/c^2) n \,$  is the rest-mass density ($c$ is the speed of light) and 
\begin{equation} 
      K =  (\Gamma - 1)  \frac{b}{(a/c^2)^\Gamma} \, .   
\label{Kappa}
\end{equation}
The energy density can thus be written as 
\begin{equation} 
   \varepsilon  =   \rho_{rm} c^2 +\frac{K}{\Gamma - 1}  \rho_{rm}^\Gamma \,.   
\label{beta-stableEOS_fit3}
\end{equation} 
The coefficients reported in Table\ \ref{tab_fit-beta-stable-eos} fit the tabular EOS for 
$\beta$-stable matter in the density range 0.08--1.30~fm$^{-3}$   
with a  ${\rm RMSRE} = 0.00018$ in the case of the N3LO$\Delta$+N2LO$\Delta1$ interaction 
and  ${\rm RMSRE} = 0.00047$ in the case of the N3LO$\Delta$+N2LO$\Delta2$ interaction.   
The curves representing these parametrized EOS in Fig.\ \ref{eos_beta} are 
indistinguishable from those obtained from the tabular EOS reported in the Appendix.

\begin{table}
\caption{Coefficients of the parametrization for the equation of state  
for $\beta$-stable nuclear matter (Eqs. (\ref{beta-stableEOS_fit1}) and (\ref{beta-stableEOS_fit2})).} 
\label{tab_fit-beta-stable-eos}
\small
\centering
\begin{tabular}{cccc}
\hline
\hline
   Model                   & $a$  & $b$ & $\Gamma$   \\
\hline
N3LO$\Delta$+N2LO$\Delta1$ &  945.199 & 293.551 & 2.82302 \\
N3LO$\Delta$+N2LO$\Delta2$ &  942.832 & 259.852 & 2.67041 \\
\hline
\end{tabular}
\tablefoot{The parameter $a$ is given in MeV, whereas the parameter $b$ is given in MeV\,fm$^{3(\Gamma-1)}$.}
\end{table}

\begin{figure}[t]
\centering
\vspace{1.1cm}
\includegraphics[width = 8cm]{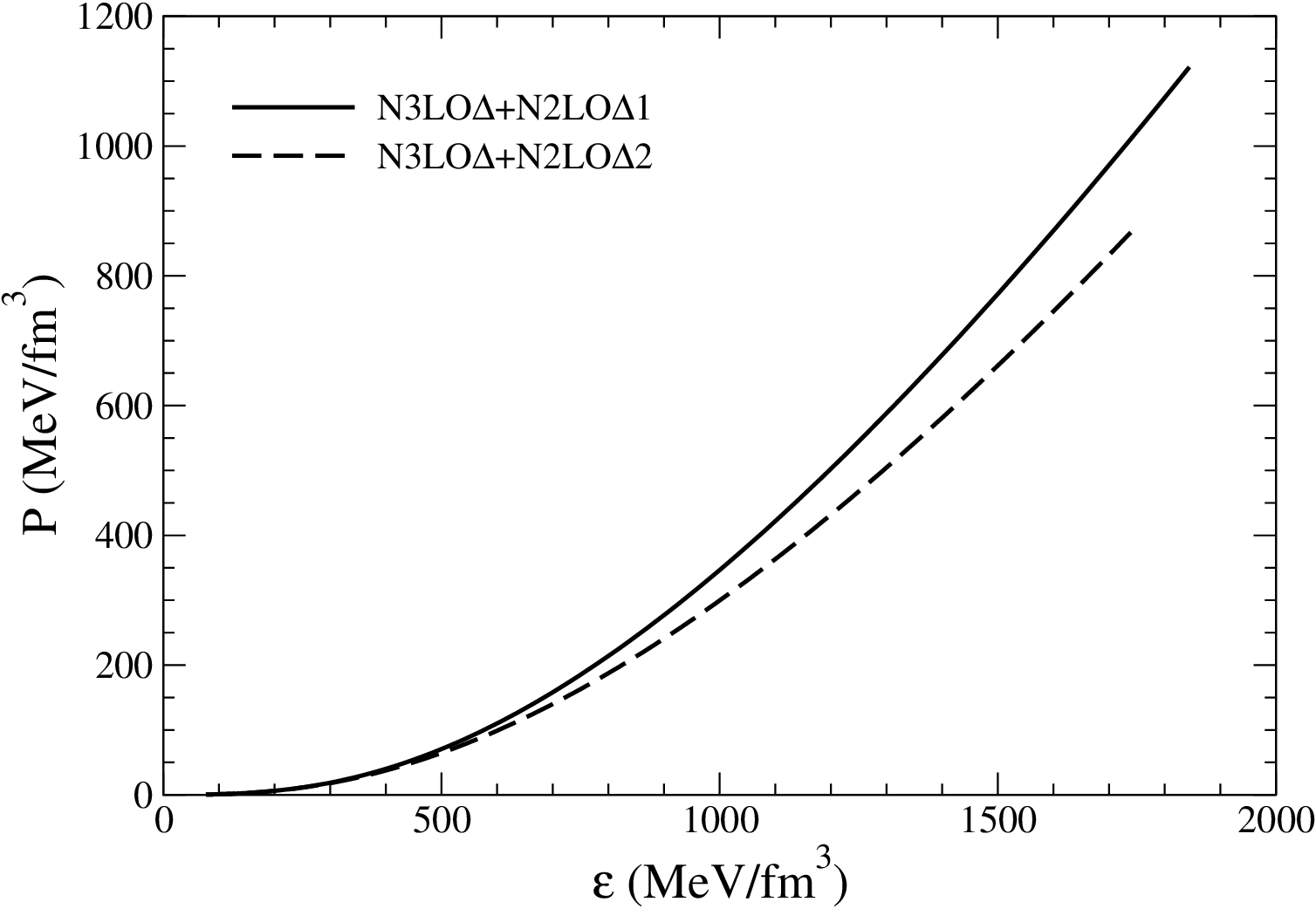}
\caption{ Equation of state of $\beta$-stable nuclear matter for the models described in the text.}
\label{eos_beta}
\end{figure}

\begin{figure}[t]
\centering
\vspace{0.8cm}
\includegraphics[width = 7.5cm]{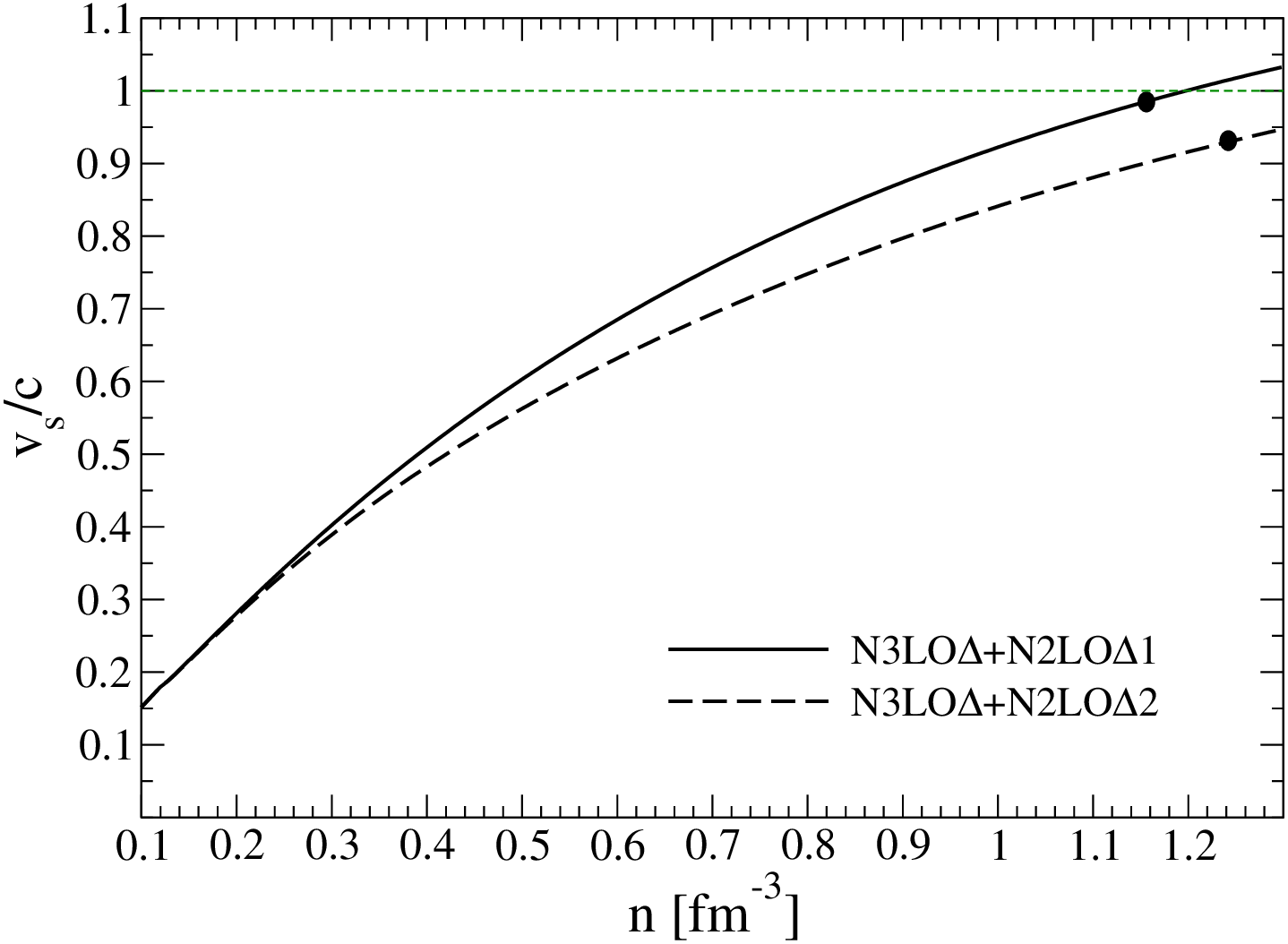}

\caption{Speed of sound $v_s/c$ (in units of the speed of light $c$) in $\beta$-stable nuclear matter  
as a function of the baryon number density $n$. 
The heavy dots on the two curves represent the central density of the neutron star maximum mass configuration 
for the corresponding EOS model.}
\label{speed-sound}
\end{figure}

In Fig.\ \ref{speed-sound} we plot the speed of sound $v_s$ in $\beta$-stable matter 
as a function of the baryon number density $n$.  
The plotted curves were obtained performing the numerical derivative of our tabular EOS  
according to the definition $v_s/c = (dP/d\varepsilon)^{1/2}$.   
Using our parametrization  (Eqs. (\ref{beta-stableEOS_fit1}) and (\ref{beta-stableEOS_fit2})) 
for the EOS of $\beta$-stable matter, we obtain  
\begin{equation} 
    \bigg(\frac{v_s}{c}\bigg)^2 = \frac{\Gamma(\Gamma-1) b n^{\Gamma-1}}{a + \Gamma b n^{\Gamma-1}} 
    = \frac{\Gamma P}{\varepsilon + P} \, .  
\label{speed-sound_fit}
\end{equation}
The corresponding curves in Fig.\ \ref{speed-sound} are indistinguishable from those obtained using 
the the numerical derivative from the tabular EOS.   
The heavy dots on both curves in Fig.\ \ref{speed-sound} represent the central density of the 
neutron star maximum mass configuration for the corresponding EOS model (see next section).  
Thus, our EOS models fulfill the causality condition $v_S/c < 1$ up to the highest densities 
reached in the corresponding neutron star configurations (see next section).

\section{Neutron star structure}

The structural properties of non-rotating neutron stars can be obtained integrating numerically 
the equation for hydrostatic equilibrium in general relativity \citep{Tol34,OV39} 
\begin{equation}
\frac{dP}{dr} = -G \, \frac{m(r)\varepsilon(r)}{c^2 r^2} \, 
                    \Bigg( 1 + \frac{P(r)}{\varepsilon(r)}\Bigg) \, 
                    \Bigg( 1 + \frac{4\pi r^3 P^{\,3}(r)}{c^2 m(r)}\Bigg) \,
                    \Bigg( 1 - \frac{2 G m(r)}{c^2 r}\Bigg)^{-1} \, 
\label{tov}
\end{equation} 
and 
\begin{equation}
     \frac{d m(r)}{d r} = \frac{4 \pi}{c^2} r^2 \varepsilon(r) \,, 
\label{enclosed-mass}
\end{equation}  
where $G$ is the gravitational constant and $m(r)$ is the gravitational mass enclosed within 
a sphere of radial coordinate $r$ (surface area $4\pi r^2$).   

Starting with a central energy density $\varepsilon_c \equiv \varepsilon(r=0)$,    
we integrate out Eqs. (\ref{tov}) and (\ref{enclosed-mass}) until the energy density   
equals the one corresponding to the density of iron $\varepsilon_{surf}/c^2 = 7.86$~g/cm$^3$.   
This condition determines the stellar surface and specifies the neutron star radius $R$  
(through the surface area $4 \pi R^2$) and the stellar gravitational mass 
\begin{equation} 
   M \equiv  m(R) = \frac{4 \pi}{c^2} \int_0^{R} dr~ r^2 \varepsilon(r). 
\label{grav-mass}
\end{equation}

The total baryon number of a star with central baryon density $n_c = n(r=0)$ 
is given by 
\begin{equation} 
  N_B  = 4 \pi  \int_0^{R} dr~ r^2 n(r)\Bigg( 1 - \frac{2 G m(r)}{c^2 r}\Bigg)^{-1/2} \,, 
\label{stellar-bar-num}
\end{equation}
and the baryonic mass (or ``rest mass'') of the neutron star is 
\begin{equation} 
  M_B = m_u N_B 
\label{baryonic-mass}
,\end{equation}
where $m_u$ is a baryonic mass unit that we take equal to 
$m_u = m(^{12}C)/12 = 1.6605 \times 10^{-24} {\rm g}$. 
Other choices for $m_u$ are sometimes used in the literature as $m_u = m_n$  or 
$m_u = m(^{56}Fe)/56$. These choices for $m_u$ only make  small changes in the calculated 
stellar binding energy since  $\Delta E_{bind}/(M_{B}c^2) \sim 0.01$.   

The total binding energy of the star is thus 
\begin{equation} 
   E_{bind} = (M_B - M) c^2 
\label{stellar-bind}
,\end{equation}
which represents the total energy liberated during the neutron star's birth.

The stellar structure equations (\ref{tov}), (\ref{enclosed-mass}), and (\ref{stellar-bar-num}) 
have been integrated using the microscopic EOS (in tabular form) for $\beta$-stable nuclear matter 
described in the previous sections to model the neutron star core, 
whereas to model the stellar crust (i.e., for nucleonic density $\le 0.08$ fm$^{-3}$)  
we have used the Baym--Pethick--Sutherland \citep{bps} and the \cite{NV73} EOS. 
The results are shown in Fig.\ \ref{MR}, where we plot the mass-radius (panel (a)) 
and mass-central density (panel (b)) relations for the considered EOS models.  
We note that our EOS models are both compatible with current measured neutron star masses and particularly 
with the mass $M = 2.01 \pm 0.04 \, M_{\odot}$ \citep{anto2013} of the neutron stars in PSR~J0348+0432. 
The hatched regions in Fig.\ \ref{MR} represent the mass--radius constraints based on   
the analysis of recent observations of both transiently accreting and bursting X-ray sources 
obtained by \cite{steiner10,steiner13b}. 
Manifestly, the neutron star configurations calculated with our EOS models are able to fulfill 
these empirical constraints on the mass--radius relationship. 

Various structural properties of the maximum mass configuration for the two considered EOS models 
are listed in Table\ \ref{tab4}. 
Our present results are in good agreement with the results of other 
calculations \citep{bbb97,apr98} based on microscopic approaches.  

The gravitational redshift of a signal emitted from the stellar surface is 
\begin{equation} 
   z_{surf} = \Bigg( 1 - \frac{2 G M}{c^2 R}\Bigg)^{-1/2} - 1 \,.  
\label{zsurf}
\end{equation} 
Thus, measurements of $z_{surf}$ of spectral lines can give  direct information on the stellar 
compactness parameter 
\begin{equation}
     x_{GR} = \frac{2 G M}{c^2 R}
\label{xgr}
\end{equation}
and consequently can place limits on the EOS for dense matter. 
The surface gravitational redshift calculated for our two EOS models is presented in  
Fig.\ \ref{redshift}. The two horizontal lines in the same figure  
represent the measured gravitational redshift  
$z = 0.35$ for the X-ray bursts source  in the low-mass X-ray binary EXO\,07482$-$676 \citep{Cottam2002} and  
$z=0.205_{-0.003}^{+0.006}$ for the isolated neutron star RX\,J0720.4$-$3125 \citep{Hambaryan2017}. 

In Fig.\ \ref{star-BE} (left panel) we plot the binding energy $E_{bind}$ versus 
the stellar gravitational mass $M$.  
Various empirical formulae have been given to describe the dependence $E_{bind}(M)$ or 
$E_{bind}(M_B)$ \citep{LY1989,prak97,Latt-Prak-2001}. 
In particular many numerical calculations have shown that there is a narrow band of possible 
values  of the stellar binding energy for a given mass implying the existence of a universal relation 
$E_{bind}(M)$ or $E_{bind}(M_B)$.  

 Our calculated binding energy for neutron stars in the mass range $1.0 M_{\odot} \leq M  \leq M_{max}$ 
can be fitted with very high accuracy using the simple relation  
\begin{equation}
     E_{bind}  = a_{bind} (M/M_{\odot})^{5/2} 
\label{BE_fit1}
\end{equation}
with $a_{bind} = 1.055 \times 10^{53}$erg  ($a_{bind} = 1.073 \times 10^{53}$erg) for the 
N3LO$\Delta$+N2LO$\Delta1$ (N3LO$\Delta$+N2LO$\Delta2$) EOS model. 

Next, in Fig.\ \ref{star-BE} (right panel) we plot the quantity $E_{bind}/(M c^2)$ as a function of 
the surface gravitational redshift $z_{surf}$. 
The results for the stellar binding energy from the numerical integration of the TOV equation 
for neutron stars in the mass range $1.0 M_{\odot} \leq M  \leq M_{max}$ 
can be fitted with very high accuracy using the simple relation 
\begin{equation}  
  \frac{E_{bind}}{M c^2} = \frac{M_B-M}{M} = t_1\, z_{surf} + t_3\, z_{surf}^3  
\label{BE_fit2}
\end{equation}
with $t_1 = 0.4505$ ($t_1 = 0.4509$) and  $t_3 = -0.4207$ ($t_3 = -0.4832$) for the 
N3LO$\Delta$+N2LO$\Delta1$  (N3LO$\Delta$+N2LO$\Delta2$) EOS model.  

The binding energy of a neutron star could be deduced from the detection of neutrinos from 
a nearby supernova. In addition, a possible measurement of $z_{surf}$ for the neutron star left behind 
by the same supernova event will give very strong constraints on the EOS of stellar matter. 
We note that  a few minutes after its birth a neutron star can be described using the EOS of cold 
(i.e., at zero temperature) and neutrino-free matter (Prakash et al. 1997).  


\begin{figure}[t]
\centering
\vspace{1.5cm}
\includegraphics[width = 8cm]{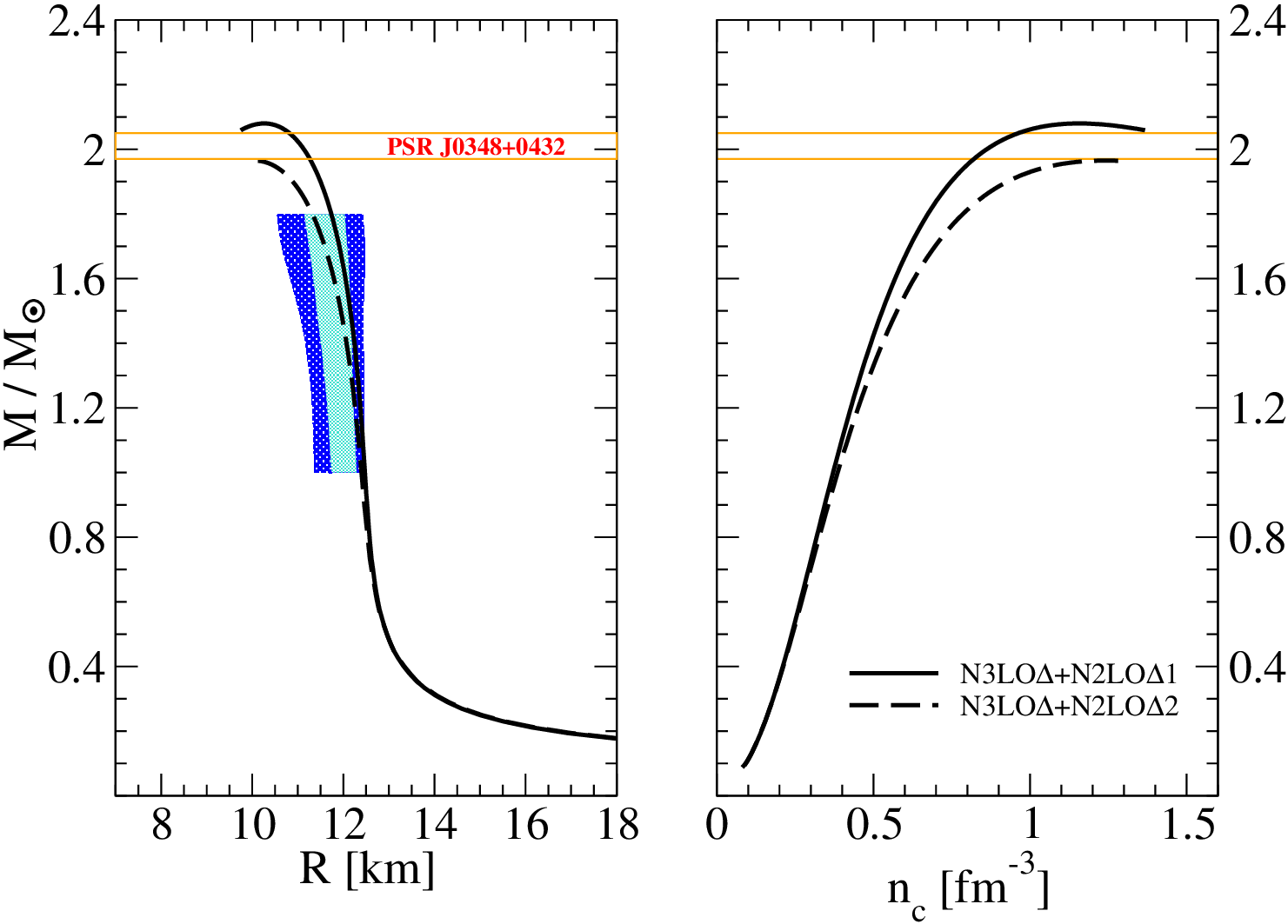}
\caption{ Mass--radius relationship (panel (a)) and  
mass--central baryon density relationship (panel (b)) 
for the two nuclear interaction models considered in this work. 
The hatched region in panel (a) represents the  mass--radius constraints obtained by  
\cite{steiner10,steiner13b}.     
The strip with boundaries marked with orange lines stands for the measured mass 
$M = 2.01 \pm 0.04 \, M_{\odot}$ \citep{anto2013} of the neutron stars in PSR~J0348+0432.}
\label{MR}
\end{figure}

\begin{table} 
\caption{Maximum mass configuration properties for the interaction models considered in this work.}  
\label{tab4}
\small
\centering
\begin{tabular}{ccccc}
\hline
\hline
Model     & $M$ ($M_\odot$) & $R$ (km) & $n_c$ (fm$^{-3}$) & $M_B$ ($M_\odot$)  \\
\hline
N3LO$\Delta$+N2LO$\Delta1$  & 2.08 & 10.26 & 1.156  & 2.45    \\ 
N3LO$\Delta$+N2LO$\Delta2$  & 1.96 & 10.04 & 1.242  & 2.29    \\ 
\hline
\end{tabular}
\tablefoot{Stellar gravitational maximum mass $M$, corresponding radius $R$,  
central baryon number density $n_c$, and baryonic maximum mass $M_B$. 
Stellar masses are given in units of the solar mass $M_\odot = 1.989\times 10^{33}$~g.}
\end{table}
%

\begin{figure}[t]
\centering
\vspace{1.5cm}
\includegraphics[width = 6.5cm]{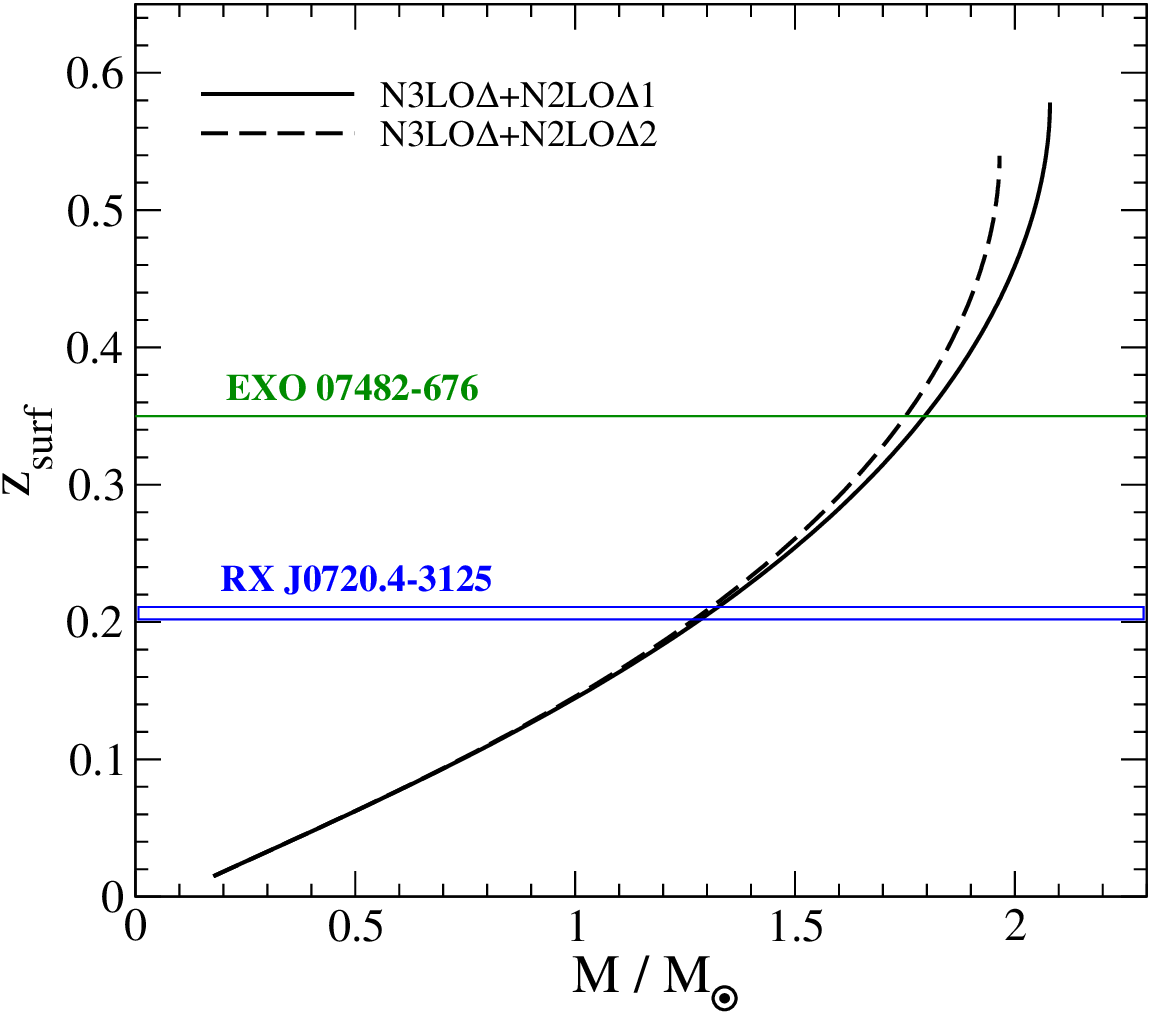}
\caption{Gravitational redshift at the neutron star surface as a function of the stellar 
gravitational mass for the two considered EOS models. 
The horizontal lines represent the measured gravitational redshift   
$z = 0.35$ for the X-ray burst source in the low-mass X-ray binary 
EXO\,07482$-$676 \citep{Cottam2002} and  $z=0.205_{-0.003}^{+0.006}$ for the 
isolated neutron star RX\,J0720.4$-$3125 \citep{Hambaryan2017}. } 
\label{redshift}
\end{figure}

\begin{figure}[t]
\centering
\vspace{1.5cm}
\includegraphics[width = 8.7cm]{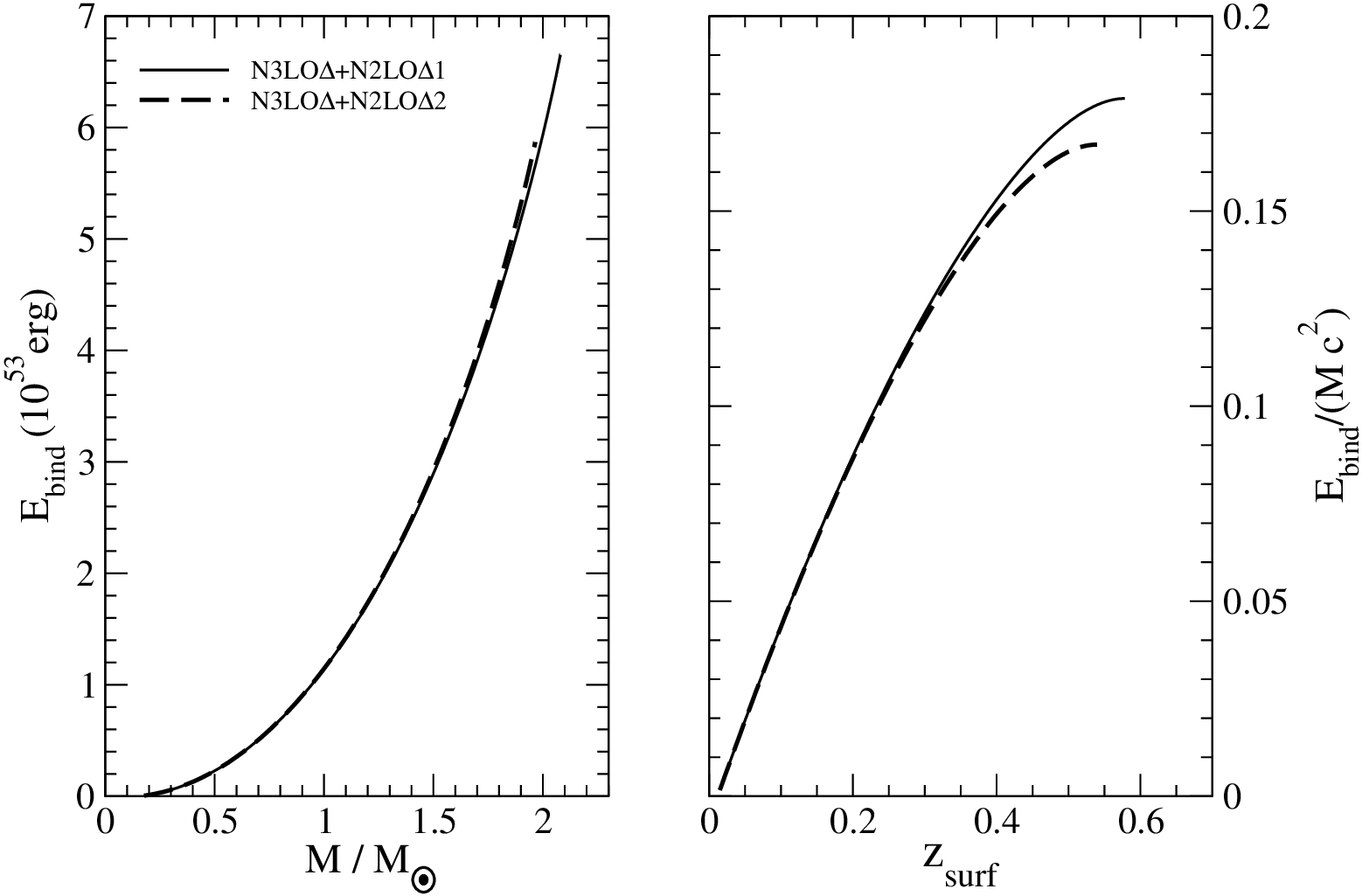}
\caption{Neutron star binding energy versus the stellar gravitational mass (left panel) and 
fractional binding energy $E_{bind}/(M c^2) = (M_B - M)/M$ as a function of the
surface gravitational redshift $z_{surf}$ (right panel) for the two considered EOS models.
} 
\label{star-BE}
\end{figure}

\section{Summary}

In this work we derived a new microscopic EOS of dense SMN, PNM, as well as asymmetric and 
$\beta$-stable nuclear matter at zero temperature using modern two-body and three-body nuclear forces 
determined in the framework of chiral perturbation theory and including the $\Delta$ isobar intermediate state.  
To this end, we employed the BHF many-body approach, which properly takes into account the short-range 
correlations arising from the strongly repulsive core in the bare NN interaction. 
This feature is particularly relevant in the case of matter at supranuclear densities.  
Our EOS models are able to reproduce the empirical saturation point of symmetric nuclear matter, 
the symmetry energy $E_{sym}$, and its slope parameter $L$ at the empirical saturation density $n_0$  
and are compatible with experimental data from collisions between heavy nuclei 
at energies ranging from a few tens of MeV up to several hundreds of MeV per nucleon.     
We used our EOS for $\beta$-stable nuclear matter to compute various structural properties   
of non-rotating neutron stars. The calculated neutron star configurations are consistent with present 
measured neutron star masses and particularly with the mass $M = 2.01 \pm 0.04 \, M_{\odot}$ of 
the neutron stars in PSR~J0348+0432.  

We provided our new EOS both in tabular form and in parametrized form ready 
to be used in numerical general relativity simulations of binary neutron star merging.    
To this purpose our zero temperature EOS needs to be supplemented by a thermal   
contribution that accounts for the sizeable increase in the internal energy at the merger. 
This is usually done by adding  an ideal-fluid component to the zero temperature EOS, which 
accounts for the shock heating \citep{rezzolla_book}.  

In a more consistent thermodynamical approach the EOS relevant for core-collapse SNe and BNS mergers 
simulations should be derived within a finite temperature many-body approach 
\citep{shen98,hempel10,steiner13a,oertel2017,togashi17}. 
We are presently working on the extension of our microscopic nuclear matter EOS  based on nuclear 
chiral interactions to the case of finite temperature for applications to numerical simulations of the 
above-mentioned astrophysical phenomena.  

In future studies we also plan to extend our present calculations to the case of  
$\beta$-stable hyperonic matter, based on nucleon-hyperon \citep{ny_chiral},  
and hyperon-hyperon \citep{yy_chiral} interactions derived in the framework of ChPT,   
and   to include hyperonic three-body interactions within this formalism.   

\section*{Acknowledgement}  
This work has been partially supported by ``NewCompstar'', COST Action MP1304.



\newpage
\appendix
\section{Tables for the equation of state of $\beta$-stable matter}
%
\tiny
\begin{longtable}{ccccc}
\tablehead{Equation of state for $\beta$-stable matter. Model N3LO$\Delta$+N2LO$\Delta1$.\\} 
\\
\hline
\hline
     $n$ (fm$^{-3}$) & $\varepsilon$ (MeV/fm$^{3}$)  & P (MeV/fm$^{3}$) & $x_p$  & $x_e$ \\
\hline
     8.0000E-02 &     7.5863E+01 &     5.1039E-01 &      3.483E-02 &      3.483E-02 \\
     8.5701E-02 &     8.1308E+01 &     5.9983E-01 &      3.590E-02 &      3.590E-02 \\
     9.1402E-02 &     8.6760E+01 &     7.0057E-01 &      3.701E-02 &      3.701E-02 \\
     9.7103E-02 &     9.2219E+01 &     8.1342E-01 &      3.815E-02 &      3.815E-02 \\
     1.0280E-01 &     9.7684E+01 &     9.3919E-01 &      3.932E-02 &      3.932E-02 \\
     1.0850E-01 &     1.0316E+02 &     1.0787E+00 &      4.051E-02 &      4.051E-02 \\
     1.1421E-01 &     1.0864E+02 &     1.2327E+00 &      4.173E-02 &      4.173E-02 \\
     1.1991E-01 &     1.1413E+02 &     1.4019E+00 &      4.296E-02 &      4.296E-02 \\
     1.2561E-01 &     1.1962E+02 &     1.5844E+00 &      4.452E-02 &      4.412E-02 \\
     1.3131E-01 &     1.2513E+02 &     1.7805E+00 &      4.635E-02 &      4.521E-02 \\
     1.3701E-01 &     1.3064E+02 &     1.9922E+00 &      4.832E-02 &      4.626E-02 \\
     1.4271E-01 &     1.3617E+02 &     2.2203E+00 &      5.036E-02 &      4.730E-02 \\
     1.4841E-01 &     1.4170E+02 &     2.4658E+00 &      5.247E-02 &      4.833E-02 \\
     1.5411E-01 &     1.4724E+02 &     2.7293E+00 &      5.462E-02 &      4.934E-02 \\
     1.5981E-01 &     1.5280E+02 &     3.0116E+00 &      5.680E-02 &      5.036E-02 \\
     1.6551E-01 &     1.5836E+02 &     3.3134E+00 &      5.901E-02 &      5.136E-02 \\
     1.7121E-01 &     1.6393E+02 &     3.6353E+00 &      6.124E-02 &      5.237E-02 \\
     1.7692E-01 &     1.6952E+02 &     3.9782E+00 &      6.347E-02 &      5.338E-02 \\
     1.8262E-01 &     1.7511E+02 &     4.3425E+00 &      6.572E-02 &      5.438E-02 \\
     1.8832E-01 &     1.8072E+02 &     4.7290E+00 &      6.798E-02 &      5.539E-02 \\
     1.9402E-01 &     1.8634E+02 &     5.1383E+00 &      7.024E-02 &      5.639E-02 \\
     1.9972E-01 &     1.9198E+02 &     5.5710E+00 &      7.250E-02 &      5.739E-02 \\
     2.0542E-01 &     1.9762E+02 &     6.0277E+00 &      7.476E-02 &      5.839E-02 \\
     2.1112E-01 &     2.0328E+02 &     6.5091E+00 &      7.703E-02 &      5.940E-02 \\
     2.1682E-01 &     2.0895E+02 &     7.0158E+00 &      7.930E-02 &      6.040E-02 \\
     2.2252E-01 &     2.1464E+02 &     7.5482E+00 &      8.156E-02 &      6.140E-02 \\
     2.2822E-01 &     2.2034E+02 &     8.1071E+00 &      8.382E-02 &      6.241E-02 \\
     2.3393E-01 &     2.2605E+02 &     8.6930E+00 &      8.608E-02 &      6.341E-02 \\
     2.3963E-01 &     2.3178E+02 &     9.3065E+00 &      8.834E-02 &      6.441E-02 \\
     2.4533E-01 &     2.3752E+02 &     9.9481E+00 &      9.059E-02 &      6.541E-02 \\
     2.5103E-01 &     2.4328E+02 &     1.0618E+01 &      9.284E-02 &      6.641E-02 \\
     2.5673E-01 &     2.4905E+02 &     1.1318E+01 &      9.508E-02 &      6.741E-02 \\
     2.6243E-01 &     2.5484E+02 &     1.2047E+01 &      9.732E-02 &      6.841E-02 \\
     2.6813E-01 &     2.6065E+02 &     1.2807E+01 &      9.955E-02 &      6.941E-02 \\
     2.7383E-01 &     2.6647E+02 &     1.3598E+01 &      1.018E-01 &      7.041E-02 \\
     2.7953E-01 &     2.7231E+02 &     1.4420E+01 &      1.040E-01 &      7.140E-02 \\
     2.8523E-01 &     2.7817E+02 &     1.5274E+01 &      1.062E-01 &      7.240E-02 \\
     2.9093E-01 &     2.8404E+02 &     1.6161E+01 &      1.084E-01 &      7.339E-02 \\
     2.9664E-01 &     2.8993E+02 &     1.7081E+01 &      1.106E-01 &      7.438E-02 \\
     3.0234E-01 &     2.9584E+02 &     1.8034E+01 &      1.128E-01 &      7.537E-02 \\
     3.0804E-01 &     3.0177E+02 &     1.9021E+01 &      1.150E-01 &      7.636E-02 \\
     3.1374E-01 &     3.0771E+02 &     2.0043E+01 &      1.172E-01 &      7.734E-02 \\
     3.1944E-01 &     3.1368E+02 &     2.1100E+01 &      1.194E-01 &      7.832E-02 \\
     3.2514E-01 &     3.1966E+02 &     2.2193E+01 &      1.215E-01 &      7.930E-02 \\
     3.3084E-01 &     3.2567E+02 &     2.3322E+01 &      1.237E-01 &      8.028E-02 \\
     3.3654E-01 &     3.3169E+02 &     2.4487E+01 &      1.258E-01 &      8.126E-02 \\
     3.4224E-01 &     3.3774E+02 &     2.5689E+01 &      1.280E-01 &      8.223E-02 \\
     3.4794E-01 &     3.4380E+02 &     2.6929E+01 &      1.301E-01 &      8.320E-02 \\
     3.5364E-01 &     3.4988E+02 &     2.8208E+01 &      1.322E-01 &      8.417E-02 \\
     3.5935E-01 &     3.5599E+02 &     2.9524E+01 &      1.343E-01 &      8.513E-02 \\
     3.6505E-01 &     3.6212E+02 &     3.0880E+01 &      1.364E-01 &      8.609E-02 \\
     3.7075E-01 &     3.6826E+02 &     3.2276E+01 &      1.385E-01 &      8.705E-02 \\
     3.7645E-01 &     3.7443E+02 &     3.3711E+01 &      1.406E-01 &      8.800E-02 \\
     3.8215E-01 &     3.8063E+02 &     3.5187E+01 &      1.427E-01 &      8.895E-02 \\
     3.8785E-01 &     3.8684E+02 &     3.6704E+01 &      1.447E-01 &      8.990E-02 \\
     3.9355E-01 &     3.9308E+02 &     3.8263E+01 &      1.468E-01 &      9.084E-02 \\
     3.9925E-01 &     3.9934E+02 &     3.9863E+01 &      1.489E-01 &      9.178E-02 \\
     4.0495E-01 &     4.0562E+02 &     4.1506E+01 &      1.509E-01 &      9.272E-02 \\
     4.1065E-01 &     4.1193E+02 &     4.3192E+01 &      1.529E-01 &      9.365E-02 \\
     4.1636E-01 &     4.1826E+02 &     4.4921E+01 &      1.549E-01 &      9.458E-02 \\
     4.2206E-01 &     4.2461E+02 &     4.6694E+01 &      1.569E-01 &      9.550E-02 \\
     4.2776E-01 &     4.3099E+02 &     4.8511E+01 &      1.589E-01 &      9.642E-02 \\
     4.3346E-01 &     4.3739E+02 &     5.0373E+01 &      1.609E-01 &      9.734E-02 \\
     4.3916E-01 &     4.4382E+02 &     5.2281E+01 &      1.629E-01 &      9.825E-02 \\
     4.4486E-01 &     4.5027E+02 &     5.4234E+01 &      1.649E-01 &      9.916E-02 \\
     4.5056E-01 &     4.5675E+02 &     5.6233E+01 &      1.668E-01 &      1.001E-01 \\
     4.5626E-01 &     4.6325E+02 &     5.8279E+01 &      1.688E-01 &      1.010E-01 \\
     4.6196E-01 &     4.6978E+02 &     6.0373E+01 &      1.707E-01 &      1.019E-01 \\
     4.6766E-01 &     4.7634E+02 &     6.2513E+01 &      1.726E-01 &      1.027E-01 \\
     4.7336E-01 &     4.8292E+02 &     6.4702E+01 &      1.745E-01 &      1.036E-01 \\
     4.7907E-01 &     4.8953E+02 &     6.6939E+01 &      1.764E-01 &      1.045E-01 \\
     4.8477E-01 &     4.9617E+02 &     6.9225E+01 &      1.783E-01 &      1.054E-01 \\
     4.9047E-01 &     5.0283E+02 &     7.1561E+01 &      1.802E-01 &      1.063E-01 \\
     4.9617E-01 &     5.0952E+02 &     7.3947E+01 &      1.821E-01 &      1.071E-01 \\
     5.0187E-01 &     5.1624E+02 &     7.6382E+01 &      1.839E-01 &      1.080E-01 \\
     5.0757E-01 &     5.2298E+02 &     7.8869E+01 &      1.858E-01 &      1.088E-01 \\
     5.1327E-01 &     5.2976E+02 &     8.1407E+01 &      1.876E-01 &      1.097E-01 \\
     5.1897E-01 &     5.3656E+02 &     8.3996E+01 &      1.894E-01 &      1.105E-01 \\
     5.2467E-01 &     5.4339E+02 &     8.6638E+01 &      1.913E-01 &      1.114E-01 \\
     5.3037E-01 &     5.5025E+02 &     8.9332E+01 &      1.931E-01 &      1.122E-01 \\
     5.3607E-01 &     5.5714E+02 &     9.2080E+01 &      1.948E-01 &      1.131E-01 \\
     5.4178E-01 &     5.6406E+02 &     9.4881E+01 &      1.966E-01 &      1.139E-01 \\
     5.4748E-01 &     5.7101E+02 &     9.7736E+01 &      1.984E-01 &      1.147E-01 \\
     5.5318E-01 &     5.7799E+02 &     1.0064E+02 &      2.002E-01 &      1.155E-01 \\
     5.5888E-01 &     5.8500E+02 &     1.0361E+02 &      2.019E-01 &      1.163E-01 \\
     5.6458E-01 &     5.9203E+02 &     1.0663E+02 &      2.036E-01 &      1.172E-01 \\
     5.7028E-01 &     5.9911E+02 &     1.0970E+02 &      2.054E-01 &      1.180E-01 \\
     5.7598E-01 &     6.0621E+02 &     1.1284E+02 &      2.071E-01 &      1.188E-01 \\
     5.8168E-01 &     6.1334E+02 &     1.1602E+02 &      2.088E-01 &      1.195E-01 \\
     5.8738E-01 &     6.2050E+02 &     1.1927E+02 &      2.105E-01 &      1.203E-01 \\
     5.9308E-01 &     6.2770E+02 &     1.2257E+02 &      2.122E-01 &      1.211E-01 \\
     5.9879E-01 &     6.3493E+02 &     1.2593E+02 &      2.138E-01 &      1.219E-01 \\
     6.0449E-01 &     6.4219E+02 &     1.2935E+02 &      2.155E-01 &      1.227E-01 \\
     6.1019E-01 &     6.4948E+02 &     1.3283E+02 &      2.171E-01 &      1.234E-01 \\
     6.1589E-01 &     6.5681E+02 &     1.3637E+02 &      2.188E-01 &      1.242E-01 \\
     6.2159E-01 &     6.6416E+02 &     1.3997E+02 &      2.204E-01 &      1.250E-01 \\
     6.2729E-01 &     6.7156E+02 &     1.4362E+02 &      2.220E-01 &      1.257E-01 \\
     6.3299E-01 &     6.7898E+02 &     1.4734E+02 &      2.236E-01 &      1.265E-01 \\
     6.3869E-01 &     6.8644E+02 &     1.5112E+02 &      2.252E-01 &      1.272E-01 \\
     6.4439E-01 &     6.9393E+02 &     1.5496E+02 &      2.268E-01 &      1.280E-01 \\
     6.5009E-01 &     7.0146E+02 &     1.5887E+02 &      2.284E-01 &      1.287E-01 \\
     6.5579E-01 &     7.0902E+02 &     1.6283E+02 &      2.299E-01 &      1.294E-01 \\
     6.6150E-01 &     7.1662E+02 &     1.6686E+02 &      2.315E-01 &      1.301E-01 \\
     6.6720E-01 &     7.2425E+02 &     1.7095E+02 &      2.330E-01 &      1.309E-01 \\
     6.7290E-01 &     7.3192E+02 &     1.7511E+02 &      2.345E-01 &      1.316E-01 \\
     6.7860E-01 &     7.3962E+02 &     1.7933E+02 &      2.361E-01 &      1.323E-01 \\
     6.8430E-01 &     7.4736E+02 &     1.8361E+02 &      2.376E-01 &      1.330E-01 \\
     6.9000E-01 &     7.5513E+02 &     1.8796E+02 &      2.391E-01 &      1.337E-01 \\
     6.9570E-01 &     7.6294E+02 &     1.9238E+02 &      2.406E-01 &      1.344E-01 \\
     7.0140E-01 &     7.7079E+02 &     1.9686E+02 &      2.420E-01 &      1.351E-01 \\
     7.0710E-01 &     7.7867E+02 &     2.0141E+02 &      2.435E-01 &      1.358E-01 \\
     7.1280E-01 &     7.8659E+02 &     2.0603E+02 &      2.450E-01 &      1.365E-01 \\
     7.1850E-01 &     7.9455E+02 &     2.1071E+02 &      2.464E-01 &      1.371E-01 \\
     7.2421E-01 &     8.0254E+02 &     2.1546E+02 &      2.478E-01 &      1.378E-01 \\
     7.2991E-01 &     8.1058E+02 &     2.2028E+02 &      2.493E-01 &      1.385E-01 \\
     7.3561E-01 &     8.1865E+02 &     2.2517E+02 &      2.507E-01 &      1.391E-01 \\
     7.4131E-01 &     8.2676E+02 &     2.3012E+02 &      2.521E-01 &      1.398E-01 \\
     7.4701E-01 &     8.3490E+02 &     2.3515E+02 &      2.535E-01 &      1.404E-01 \\
     7.5271E-01 &     8.4309E+02 &     2.4025E+02 &      2.549E-01 &      1.411E-01 \\
     7.5841E-01 &     8.5131E+02 &     2.4541E+02 &      2.563E-01 &      1.417E-01 \\
     7.6411E-01 &     8.5958E+02 &     2.5065E+02 &      2.576E-01 &      1.424E-01 \\
     7.6981E-01 &     8.6788E+02 &     2.5596E+02 &      2.590E-01 &      1.430E-01 \\
     7.7551E-01 &     8.7622E+02 &     2.6134E+02 &      2.603E-01 &      1.436E-01 \\
     7.8121E-01 &     8.8461E+02 &     2.6679E+02 &      2.617E-01 &      1.443E-01 \\
     7.8692E-01 &     8.9303E+02 &     2.7232E+02 &      2.630E-01 &      1.449E-01 \\
     7.9262E-01 &     9.0149E+02 &     2.7792E+02 &      2.643E-01 &      1.455E-01 \\
     7.9832E-01 &     9.0999E+02 &     2.8359E+02 &      2.656E-01 &      1.461E-01 \\
     8.0402E-01 &     9.1854E+02 &     2.8934E+02 &      2.669E-01 &      1.467E-01 \\
     8.0972E-01 &     9.2712E+02 &     2.9516E+02 &      2.682E-01 &      1.473E-01 \\
     8.1542E-01 &     9.3575E+02 &     3.0106E+02 &      2.695E-01 &      1.479E-01 \\
     8.2112E-01 &     9.4442E+02 &     3.0703E+02 &      2.708E-01 &      1.485E-01 \\
     8.2682E-01 &     9.5313E+02 &     3.1308E+02 &      2.720E-01 &      1.491E-01 \\
     8.3252E-01 &     9.6188E+02 &     3.1920E+02 &      2.733E-01 &      1.497E-01 \\
     8.3822E-01 &     9.7067E+02 &     3.2540E+02 &      2.745E-01 &      1.503E-01 \\
     8.4393E-01 &     9.7951E+02 &     3.3168E+02 &      2.757E-01 &      1.509E-01 \\
     8.4963E-01 &     9.8839E+02 &     3.3803E+02 &      2.770E-01 &      1.514E-01 \\
     8.5533E-01 &     9.9731E+02 &     3.4447E+02 &      2.782E-01 &      1.520E-01 \\
     8.6103E-01 &     1.0063E+03 &     3.5098E+02 &      2.794E-01 &      1.526E-01 \\
     8.6673E-01 &     1.0153E+03 &     3.5757E+02 &      2.806E-01 &      1.531E-01 \\
     8.7243E-01 &     1.0243E+03 &     3.6424E+02 &      2.818E-01 &      1.537E-01 \\
     8.7813E-01 &     1.0334E+03 &     3.7098E+02 &      2.830E-01 &      1.542E-01 \\
     8.8383E-01 &     1.0426E+03 &     3.7781E+02 &      2.841E-01 &      1.548E-01 \\
     8.8953E-01 &     1.0518E+03 &     3.8472E+02 &      2.853E-01 &      1.553E-01 \\
     8.9523E-01 &     1.0610E+03 &     3.9171E+02 &      2.865E-01 &      1.559E-01 \\
     9.0093E-01 &     1.0703E+03 &     3.9879E+02 &      2.876E-01 &      1.564E-01 \\
     9.0664E-01 &     1.0796E+03 &     4.0594E+02 &      2.888E-01 &      1.570E-01 \\
     9.1234E-01 &     1.0889E+03 &     4.1318E+02 &      2.899E-01 &      1.575E-01 \\
     9.1804E-01 &     1.0983E+03 &     4.2050E+02 &      2.910E-01 &      1.580E-01 \\
     9.2374E-01 &     1.1078E+03 &     4.2790E+02 &      2.921E-01 &      1.585E-01 \\
     9.2944E-01 &     1.1173E+03 &     4.3538E+02 &      2.932E-01 &      1.591E-01 \\
     9.3514E-01 &     1.1268E+03 &     4.4295E+02 &      2.943E-01 &      1.596E-01 \\
     9.4084E-01 &     1.1364E+03 &     4.5061E+02 &      2.954E-01 &      1.601E-01 \\
     9.4654E-01 &     1.1461E+03 &     4.5835E+02 &      2.965E-01 &      1.606E-01 \\
     9.5224E-01 &     1.1558E+03 &     4.6617E+02 &      2.976E-01 &      1.611E-01 \\
     9.5794E-01 &     1.1655E+03 &     4.7408E+02 &      2.987E-01 &      1.616E-01 \\
     9.6364E-01 &     1.1753E+03 &     4.8208E+02 &      2.997E-01 &      1.621E-01 \\
     9.6935E-01 &     1.1851E+03 &     4.9016E+02 &      3.008E-01 &      1.626E-01 \\
     9.7505E-01 &     1.1950E+03 &     4.9833E+02 &      3.018E-01 &      1.631E-01 \\
     9.8075E-01 &     1.2049E+03 &     5.0659E+02 &      3.029E-01 &      1.636E-01 \\
     9.8645E-01 &     1.2149E+03 &     5.1493E+02 &      3.039E-01 &      1.640E-01 \\
     9.9215E-01 &     1.2249E+03 &     5.2337E+02 &      3.049E-01 &      1.645E-01 \\
     9.9785E-01 &     1.2350E+03 &     5.3189E+02 &      3.059E-01 &      1.650E-01 \\
     1.0036E+00 &     1.2451E+03 &     5.4050E+02 &      3.069E-01 &      1.655E-01 \\
     1.0093E+00 &     1.2553E+03 &     5.4920E+02 &      3.079E-01 &      1.659E-01 \\
     1.0150E+00 &     1.2655E+03 &     5.5800E+02 &      3.089E-01 &      1.664E-01 \\
     1.0207E+00 &     1.2757E+03 &     5.6688E+02 &      3.099E-01 &      1.669E-01 \\
     1.0264E+00 &     1.2861E+03 &     5.7585E+02 &      3.109E-01 &      1.673E-01 \\
     1.0321E+00 &     1.2964E+03 &     5.8491E+02 &      3.119E-01 &      1.678E-01 \\
     1.0378E+00 &     1.3069E+03 &     5.9407E+02 &      3.128E-01 &      1.682E-01 \\
     1.0435E+00 &     1.3173E+03 &     6.0332E+02 &      3.138E-01 &      1.687E-01 \\
     1.0492E+00 &     1.3278E+03 &     6.1266E+02 &      3.147E-01 &      1.691E-01 \\
     1.0549E+00 &     1.3384E+03 &     6.2209E+02 &      3.157E-01 &      1.696E-01 \\
     1.0606E+00 &     1.3490E+03 &     6.3162E+02 &      3.166E-01 &      1.700E-01 \\
     1.0663E+00 &     1.3597E+03 &     6.4124E+02 &      3.175E-01 &      1.704E-01 \\
     1.0720E+00 &     1.3704E+03 &     6.5095E+02 &      3.185E-01 &      1.709E-01 \\
     1.0777E+00 &     1.3812E+03 &     6.6076E+02 &      3.194E-01 &      1.713E-01 \\
     1.0834E+00 &     1.3920E+03 &     6.7067E+02 &      3.203E-01 &      1.717E-01 \\
     1.0891E+00 &     1.4029E+03 &     6.8067E+02 &      3.212E-01 &      1.721E-01 \\
     1.0948E+00 &     1.4138E+03 &     6.9076E+02 &      3.221E-01 &      1.726E-01 \\
     1.1005E+00 &     1.4248E+03 &     7.0096E+02 &      3.230E-01 &      1.730E-01 \\
     1.1062E+00 &     1.4359E+03 &     7.1124E+02 &      3.239E-01 &      1.734E-01 \\
     1.1119E+00 &     1.4470E+03 &     7.2163E+02 &      3.248E-01 &      1.738E-01 \\
     1.1176E+00 &     1.4581E+03 &     7.3211E+02 &      3.256E-01 &      1.742E-01 \\
     1.1233E+00 &     1.4693E+03 &     7.4270E+02 &      3.265E-01 &      1.746E-01 \\
     1.1290E+00 &     1.4806E+03 &     7.5338E+02 &      3.274E-01 &      1.750E-01 \\
     1.1347E+00 &     1.4919E+03 &     7.6416E+02 &      3.282E-01 &      1.754E-01 \\
     1.1404E+00 &     1.5032E+03 &     7.7503E+02 &      3.291E-01 &      1.758E-01 \\
     1.1461E+00 &     1.5146E+03 &     7.8601E+02 &      3.299E-01 &      1.762E-01 \\
     1.1518E+00 &     1.5261E+03 &     7.9709E+02 &      3.307E-01 &      1.766E-01 \\
     1.1575E+00 &     1.5376E+03 &     8.0827E+02 &      3.316E-01 &      1.770E-01 \\
     1.1632E+00 &     1.5492E+03 &     8.1955E+02 &      3.324E-01 &      1.774E-01 \\
     1.1689E+00 &     1.5609E+03 &     8.3093E+02 &      3.332E-01 &      1.777E-01 \\
     1.1746E+00 &     1.5726E+03 &     8.4241E+02 &      3.340E-01 &      1.781E-01 \\
     1.1803E+00 &     1.5843E+03 &     8.5400E+02 &      3.348E-01 &      1.785E-01 \\
     1.1860E+00 &     1.5961E+03 &     8.6568E+02 &      3.356E-01 &      1.789E-01 \\
     1.1917E+00 &     1.6080E+03 &     8.7748E+02 &      3.364E-01 &      1.793E-01 \\
     1.1974E+00 &     1.6199E+03 &     8.8937E+02 &      3.372E-01 &      1.796E-01 \\
     1.2031E+00 &     1.6319E+03 &     9.0137E+02 &      3.380E-01 &      1.800E-01 \\
     1.2088E+00 &     1.6439E+03 &     9.1347E+02 &      3.388E-01 &      1.804E-01 \\
     1.2145E+00 &     1.6560E+03 &     9.2568E+02 &      3.396E-01 &      1.807E-01 \\
     1.2202E+00 &     1.6681E+03 &     9.3799E+02 &      3.403E-01 &      1.811E-01 \\
     1.2259E+00 &     1.6803E+03 &     9.5040E+02 &      3.411E-01 &      1.814E-01 \\
     1.2316E+00 &     1.6926E+03 &     9.6293E+02 &      3.419E-01 &      1.818E-01 \\
     1.2373E+00 &     1.7049E+03 &     9.7556E+02 &      3.426E-01 &      1.821E-01 \\
     1.2430E+00 &     1.7173E+03 &     9.8829E+02 &      3.434E-01 &      1.825E-01 \\
     1.2487E+00 &     1.7297E+03 &     1.0011E+03 &      3.441E-01 &      1.828E-01 \\
     1.2544E+00 &     1.7422E+03 &     1.0141E+03 &      3.448E-01 &      1.832E-01 \\
     1.2601E+00 &     1.7548E+03 &     1.0271E+03 &      3.456E-01 &      1.835E-01 \\
     1.2658E+00 &     1.7674E+03 &     1.0403E+03 &      3.463E-01 &      1.839E-01 \\
     1.2715E+00 &     1.7801E+03 &     1.0536E+03 &      3.470E-01 &      1.842E-01 \\
     1.2772E+00 &     1.7928E+03 &     1.0670E+03 &      3.478E-01 &      1.845E-01 \\
     1.2829E+00 &     1.8056E+03 &     1.0805E+03 &      3.485E-01 &      1.849E-01 \\
     1.2886E+00 &     1.8185E+03 &     1.0941E+03 &      3.492E-01 &      1.852E-01 \\
     1.2943E+00 &     1.8314E+03 &     1.1078E+03 &      3.499E-01 &      1.855E-01 \\
     1.3000E+00 &     1.8444E+03 &     1.1216E+03 &      3.506E-01 &      1.858E-01 \\
\hline
\end{longtable} 
\tablefoot{The different entries in the table from Col. 1 to Col. 5 are respectively 
the baryon number density $n$, the total energy density $\varepsilon$,  
the total pressure $P$, the proton fraction $x_p$, and the  electron fraction $x_e$. 
The muon fraction is given by $x_\mu = x_p - x_e$. }   
\label{tab_A1}

%
%
\tiny
\begin{longtable}{ccccc}
\tablehead{Equation of state for $\beta$-stable matter. Model N3LO$\Delta$+N2LO$\Delta2$. \\} 
\\
\hline
\hline
     $n$ (fm$^{-3}$) & $\varepsilon$ (MeV/fm$^{3}$)  & P (MeV/fm$^{3}$) & $x_p$  & $x_e$ \\
\hline
     8.0000E-02 &     7.5865E+01 &     5.1139E-01 &      3.403E-02 &      3.403E-02 \\
     8.5701E-02 &     8.1311E+01 &     6.0107E-01 &      3.512E-02 &      3.512E-02 \\
     9.1402E-02 &     8.6763E+01 &     7.0203E-01 &      3.624E-02 &      3.624E-02 \\
     9.7103E-02 &     9.2222E+01 &     8.1505E-01 &      3.738E-02 &      3.738E-02 \\
     1.0280E-01 &     9.7688E+01 &     9.4093E-01 &      3.855E-02 &      3.855E-02 \\
     1.0850E-01 &     1.0316E+02 &     1.0804E+00 &      3.975E-02 &      3.975E-02 \\
     1.1421E-01 &     1.0864E+02 &     1.2343E+00 &      4.098E-02 &      4.098E-02 \\
     1.1991E-01 &     1.1413E+02 &     1.4033E+00 &      4.223E-02 &      4.223E-02 \\
     1.2561E-01 &     1.1963E+02 &     1.5862E+00 &      4.370E-02 &      4.343E-02 \\
     1.3131E-01 &     1.2513E+02 &     1.7821E+00 &      4.551E-02 &      4.456E-02 \\
     1.3701E-01 &     1.3065E+02 &     1.9930E+00 &      4.748E-02 &      4.564E-02 \\
     1.4271E-01 &     1.3617E+02 &     2.2198E+00 &      4.955E-02 &      4.672E-02 \\
     1.4841E-01 &     1.4171E+02 &     2.4635E+00 &      5.169E-02 &      4.778E-02 \\
     1.5411E-01 &     1.4725E+02 &     2.7246E+00 &      5.389E-02 &      4.884E-02 \\
     1.5981E-01 &     1.5280E+02 &     3.0039E+00 &      5.612E-02 &      4.990E-02 \\
     1.6551E-01 &     1.5836E+02 &     3.3018E+00 &      5.839E-02 &      5.096E-02 \\
     1.7121E-01 &     1.6394E+02 &     3.6191E+00 &      6.069E-02 &      5.202E-02 \\
     1.7692E-01 &     1.6952E+02 &     3.9563E+00 &      6.301E-02 &      5.308E-02 \\
     1.8262E-01 &     1.7512E+02 &     4.3139E+00 &      6.535E-02 &      5.415E-02 \\
     1.8832E-01 &     1.8073E+02 &     4.6924E+00 &      6.771E-02 &      5.522E-02 \\
     1.9402E-01 &     1.8634E+02 &     5.0924E+00 &      7.008E-02 &      5.629E-02 \\
     1.9972E-01 &     1.9198E+02 &     5.5144E+00 &      7.246E-02 &      5.737E-02 \\
     2.0542E-01 &     1.9762E+02 &     5.9588E+00 &      7.485E-02 &      5.845E-02 \\
     2.1112E-01 &     2.0328E+02 &     6.4262E+00 &      7.726E-02 &      5.953E-02 \\
     2.1682E-01 &     2.0894E+02 &     6.9169E+00 &      7.967E-02 &      6.062E-02 \\
     2.2252E-01 &     2.1463E+02 &     7.4315E+00 &      8.209E-02 &      6.172E-02 \\
     2.2822E-01 &     2.2032E+02 &     7.9704E+00 &      8.451E-02 &      6.281E-02 \\
     2.3393E-01 &     2.2603E+02 &     8.5339E+00 &      8.694E-02 &      6.391E-02 \\
     2.3963E-01 &     2.3176E+02 &     9.1225E+00 &      8.938E-02 &      6.502E-02 \\
     2.4533E-01 &     2.3749E+02 &     9.7367E+00 &      9.182E-02 &      6.612E-02 \\
     2.5103E-01 &     2.4325E+02 &     1.0377E+01 &      9.426E-02 &      6.723E-02 \\
     2.5673E-01 &     2.4901E+02 &     1.1043E+01 &      9.670E-02 &      6.834E-02 \\
     2.6243E-01 &     2.5480E+02 &     1.1736E+01 &      9.915E-02 &      6.946E-02 \\
     2.6813E-01 &     2.6059E+02 &     1.2456E+01 &      1.016E-01 &      7.057E-02 \\
     2.7383E-01 &     2.6641E+02 &     1.3204E+01 &      1.040E-01 &      7.169E-02 \\
     2.7953E-01 &     2.7224E+02 &     1.3979E+01 &      1.065E-01 &      7.281E-02 \\
     2.8523E-01 &     2.7808E+02 &     1.4782E+01 &      1.089E-01 &      7.393E-02 \\
     2.9093E-01 &     2.8394E+02 &     1.5614E+01 &      1.114E-01 &      7.506E-02 \\
     2.9664E-01 &     2.8982E+02 &     1.6474E+01 &      1.138E-01 &      7.618E-02 \\
     3.0234E-01 &     2.9572E+02 &     1.7364E+01 &      1.163E-01 &      7.730E-02 \\
     3.0804E-01 &     3.0163E+02 &     1.8283E+01 &      1.187E-01 &      7.843E-02 \\
     3.1374E-01 &     3.0756E+02 &     1.9231E+01 &      1.212E-01 &      7.955E-02 \\
     3.1944E-01 &     3.1351E+02 &     2.0210E+01 &      1.236E-01 &      8.067E-02 \\
     3.2514E-01 &     3.1947E+02 &     2.1219E+01 &      1.260E-01 &      8.180E-02 \\
     3.3084E-01 &     3.2545E+02 &     2.2259E+01 &      1.285E-01 &      8.292E-02 \\
     3.3654E-01 &     3.3145E+02 &     2.3330E+01 &      1.309E-01 &      8.404E-02 \\
     3.4224E-01 &     3.3747E+02 &     2.4432E+01 &      1.333E-01 &      8.516E-02 \\
     3.4794E-01 &     3.4351E+02 &     2.5565E+01 &      1.357E-01 &      8.628E-02 \\
     3.5364E-01 &     3.4957E+02 &     2.6730E+01 &      1.381E-01 &      8.740E-02 \\
     3.5935E-01 &     3.5564E+02 &     2.7928E+01 &      1.405E-01 &      8.851E-02 \\
     3.6505E-01 &     3.6174E+02 &     2.9157E+01 &      1.429E-01 &      8.962E-02 \\
     3.7075E-01 &     3.6785E+02 &     3.0420E+01 &      1.453E-01 &      9.073E-02 \\
     3.7645E-01 &     3.7399E+02 &     3.1715E+01 &      1.477E-01 &      9.184E-02 \\
     3.8215E-01 &     3.8014E+02 &     3.3043E+01 &      1.501E-01 &      9.295E-02 \\
     3.8785E-01 &     3.8631E+02 &     3.4405E+01 &      1.524E-01 &      9.405E-02 \\
     3.9355E-01 &     3.9251E+02 &     3.5801E+01 &      1.548E-01 &      9.515E-02 \\
     3.9925E-01 &     3.9872E+02 &     3.7231E+01 &      1.571E-01 &      9.624E-02 \\
     4.0495E-01 &     4.0496E+02 &     3.8695E+01 &      1.595E-01 &      9.733E-02 \\
     4.1065E-01 &     4.1121E+02 &     4.0194E+01 &      1.618E-01 &      9.842E-02 \\
     4.1636E-01 &     4.1749E+02 &     4.1727E+01 &      1.641E-01 &      9.950E-02 \\
     4.2206E-01 &     4.2379E+02 &     4.3296E+01 &      1.664E-01 &      1.006E-01 \\
     4.2776E-01 &     4.3011E+02 &     4.4900E+01 &      1.687E-01 &      1.017E-01 \\
     4.3346E-01 &     4.3645E+02 &     4.6539E+01 &      1.710E-01 &      1.027E-01 \\
     4.3916E-01 &     4.4281E+02 &     4.8215E+01 &      1.733E-01 &      1.038E-01 \\
     4.4486E-01 &     4.4920E+02 &     4.9926E+01 &      1.755E-01 &      1.048E-01 \\
     4.5056E-01 &     4.5561E+02 &     5.1674E+01 &      1.778E-01 &      1.059E-01 \\
     4.5626E-01 &     4.6204E+02 &     5.3459E+01 &      1.800E-01 &      1.070E-01 \\
     4.6196E-01 &     4.6849E+02 &     5.5281E+01 &      1.823E-01 &      1.080E-01 \\
     4.6766E-01 &     4.7496E+02 &     5.7140E+01 &      1.845E-01 &      1.090E-01 \\
     4.7336E-01 &     4.8146E+02 &     5.9036E+01 &      1.867E-01 &      1.101E-01 \\
     4.7907E-01 &     4.8798E+02 &     6.0970E+01 &      1.889E-01 &      1.111E-01 \\
     4.8477E-01 &     4.9453E+02 &     6.2942E+01 &      1.910E-01 &      1.121E-01 \\
     4.9047E-01 &     5.0110E+02 &     6.4953E+01 &      1.932E-01 &      1.131E-01 \\
     4.9617E-01 &     5.0769E+02 &     6.7002E+01 &      1.954E-01 &      1.142E-01 \\
     5.0187E-01 &     5.1430E+02 &     6.9090E+01 &      1.975E-01 &      1.152E-01 \\
     5.0757E-01 &     5.2094E+02 &     7.1217E+01 &      1.996E-01 &      1.162E-01 \\
     5.1327E-01 &     5.2760E+02 &     7.3383E+01 &      2.017E-01 &      1.172E-01 \\
     5.1897E-01 &     5.3429E+02 &     7.5589E+01 &      2.038E-01 &      1.181E-01 \\
     5.2467E-01 &     5.4100E+02 &     7.7835E+01 &      2.059E-01 &      1.191E-01 \\
     5.3037E-01 &     5.4774E+02 &     8.0120E+01 &      2.080E-01 &      1.201E-01 \\
     5.3607E-01 &     5.5450E+02 &     8.2447E+01 &      2.101E-01 &      1.211E-01 \\
     5.4178E-01 &     5.6129E+02 &     8.4814E+01 &      2.121E-01 &      1.220E-01 \\
     5.4748E-01 &     5.6810E+02 &     8.7222E+01 &      2.141E-01 &      1.230E-01 \\
     5.5318E-01 &     5.7494E+02 &     8.9671E+01 &      2.162E-01 &      1.239E-01 \\
     5.5888E-01 &     5.8180E+02 &     9.2162E+01 &      2.182E-01 &      1.249E-01 \\
     5.6458E-01 &     5.8868E+02 &     9.4695E+01 &      2.202E-01 &      1.258E-01 \\
     5.7028E-01 &     5.9560E+02 &     9.7270E+01 &      2.221E-01 &      1.267E-01 \\
     5.7598E-01 &     6.0254E+02 &     9.9887E+01 &      2.241E-01 &      1.277E-01 \\
     5.8168E-01 &     6.0950E+02 &     1.0255E+02 &      2.260E-01 &      1.286E-01 \\
     5.8738E-01 &     6.1650E+02 &     1.0525E+02 &      2.280E-01 &      1.295E-01 \\
     5.9308E-01 &     6.2351E+02 &     1.0800E+02 &      2.299E-01 &      1.304E-01 \\
     5.9879E-01 &     6.3056E+02 &     1.1078E+02 &      2.318E-01 &      1.313E-01 \\
     6.0449E-01 &     6.3763E+02 &     1.1362E+02 &      2.337E-01 &      1.322E-01 \\
     6.1019E-01 &     6.4473E+02 &     1.1649E+02 &      2.356E-01 &      1.331E-01 \\
     6.1589E-01 &     6.5185E+02 &     1.1942E+02 &      2.374E-01 &      1.339E-01 \\
     6.2159E-01 &     6.5901E+02 &     1.2238E+02 &      2.393E-01 &      1.348E-01 \\
     6.2729E-01 &     6.6619E+02 &     1.2539E+02 &      2.411E-01 &      1.357E-01 \\
     6.3299E-01 &     6.7340E+02 &     1.2845E+02 &      2.429E-01 &      1.365E-01 \\
     6.3869E-01 &     6.8063E+02 &     1.3155E+02 &      2.447E-01 &      1.374E-01 \\
     6.4439E-01 &     6.8789E+02 &     1.3470E+02 &      2.465E-01 &      1.382E-01 \\
     6.5009E-01 &     6.9519E+02 &     1.3789E+02 &      2.483E-01 &      1.391E-01 \\
     6.5579E-01 &     7.0251E+02 &     1.4113E+02 &      2.501E-01 &      1.399E-01 \\
     6.6150E-01 &     7.0985E+02 &     1.4442E+02 &      2.518E-01 &      1.407E-01 \\
     6.6720E-01 &     7.1723E+02 &     1.4775E+02 &      2.536E-01 &      1.415E-01 \\
     6.7290E-01 &     7.2464E+02 &     1.5113E+02 &      2.553E-01 &      1.424E-01 \\
     6.7860E-01 &     7.3207E+02 &     1.5456E+02 &      2.570E-01 &      1.432E-01 \\
     6.8430E-01 &     7.3953E+02 &     1.5803E+02 &      2.587E-01 &      1.440E-01 \\
     6.9000E-01 &     7.4702E+02 &     1.6156E+02 &      2.604E-01 &      1.447E-01 \\
     6.9570E-01 &     7.5455E+02 &     1.6513E+02 &      2.620E-01 &      1.455E-01 \\
     7.0140E-01 &     7.6210E+02 &     1.6875E+02 &      2.637E-01 &      1.463E-01 \\
     7.0710E-01 &     7.6968E+02 &     1.7242E+02 &      2.653E-01 &      1.471E-01 \\
     7.1280E-01 &     7.7729E+02 &     1.7614E+02 &      2.669E-01 &      1.478E-01 \\
     7.1850E-01 &     7.8493E+02 &     1.7990E+02 &      2.685E-01 &      1.486E-01 \\
     7.2421E-01 &     7.9260E+02 &     1.8372E+02 &      2.701E-01 &      1.494E-01 \\
     7.2991E-01 &     8.0030E+02 &     1.8759E+02 &      2.717E-01 &      1.501E-01 \\
     7.3561E-01 &     8.0803E+02 &     1.9151E+02 &      2.733E-01 &      1.508E-01 \\
     7.4131E-01 &     8.1579E+02 &     1.9548E+02 &      2.749E-01 &      1.516E-01 \\
     7.4701E-01 &     8.2359E+02 &     1.9950E+02 &      2.764E-01 &      1.523E-01 \\
     7.5271E-01 &     8.3141E+02 &     2.0357E+02 &      2.779E-01 &      1.530E-01 \\
     7.5841E-01 &     8.3926E+02 &     2.0769E+02 &      2.794E-01 &      1.537E-01 \\
     7.6411E-01 &     8.4715E+02 &     2.1186E+02 &      2.810E-01 &      1.544E-01 \\
     7.6981E-01 &     8.5507E+02 &     2.1609E+02 &      2.824E-01 &      1.551E-01 \\
     7.7551E-01 &     8.6301E+02 &     2.2037E+02 &      2.839E-01 &      1.558E-01 \\
     7.8121E-01 &     8.7099E+02 &     2.2470E+02 &      2.854E-01 &      1.565E-01 \\
     7.8692E-01 &     8.7901E+02 &     2.2908E+02 &      2.868E-01 &      1.572E-01 \\
     7.9262E-01 &     8.8705E+02 &     2.3352E+02 &      2.883E-01 &      1.579E-01 \\
     7.9832E-01 &     8.9513E+02 &     2.3801E+02 &      2.897E-01 &      1.586E-01 \\
     8.0402E-01 &     9.0323E+02 &     2.4255E+02 &      2.911E-01 &      1.592E-01 \\
     8.0972E-01 &     9.1137E+02 &     2.4715E+02 &      2.925E-01 &      1.599E-01 \\
     8.1542E-01 &     9.1955E+02 &     2.5181E+02 &      2.939E-01 &      1.605E-01 \\
     8.2112E-01 &     9.2775E+02 &     2.5651E+02 &      2.953E-01 &      1.612E-01 \\
     8.2682E-01 &     9.3599E+02 &     2.6128E+02 &      2.967E-01 &      1.618E-01 \\
     8.3252E-01 &     9.4426E+02 &     2.6609E+02 &      2.980E-01 &      1.625E-01 \\
     8.3822E-01 &     9.5257E+02 &     2.7097E+02 &      2.994E-01 &      1.631E-01 \\
     8.4393E-01 &     9.6091E+02 &     2.7590E+02 &      3.007E-01 &      1.637E-01 \\
     8.4963E-01 &     9.6928E+02 &     2.8088E+02 &      3.020E-01 &      1.644E-01 \\
     8.5533E-01 &     9.7768E+02 &     2.8592E+02 &      3.034E-01 &      1.650E-01 \\
     8.6103E-01 &     9.8612E+02 &     2.9102E+02 &      3.047E-01 &      1.656E-01 \\
     8.6673E-01 &     9.9460E+02 &     2.9618E+02 &      3.059E-01 &      1.662E-01 \\
     8.7243E-01 &     1.0031E+03 &     3.0139E+02 &      3.072E-01 &      1.668E-01 \\
     8.7813E-01 &     1.0116E+03 &     3.0666E+02 &      3.085E-01 &      1.674E-01 \\
     8.8383E-01 &     1.0202E+03 &     3.1199E+02 &      3.097E-01 &      1.680E-01 \\
     8.8953E-01 &     1.0288E+03 &     3.1738E+02 &      3.110E-01 &      1.685E-01 \\
     8.9523E-01 &     1.0375E+03 &     3.2282E+02 &      3.122E-01 &      1.691E-01 \\
     9.0093E-01 &     1.0462E+03 &     3.2833E+02 &      3.134E-01 &      1.697E-01 \\
     9.0664E-01 &     1.0549E+03 &     3.3389E+02 &      3.147E-01 &      1.703E-01 \\
     9.1234E-01 &     1.0636E+03 &     3.3951E+02 &      3.159E-01 &      1.708E-01 \\
     9.1804E-01 &     1.0724E+03 &     3.4519E+02 &      3.170E-01 &      1.714E-01 \\
     9.2374E-01 &     1.0812E+03 &     3.5093E+02 &      3.182E-01 &      1.719E-01 \\
     9.2944E-01 &     1.0901E+03 &     3.5673E+02 &      3.194E-01 &      1.725E-01 \\
     9.3514E-01 &     1.0990E+03 &     3.6260E+02 &      3.206E-01 &      1.730E-01 \\
     9.4084E-01 &     1.1079E+03 &     3.6852E+02 &      3.217E-01 &      1.736E-01 \\
     9.4654E-01 &     1.1169E+03 &     3.7450E+02 &      3.228E-01 &      1.741E-01 \\
     9.5224E-01 &     1.1259E+03 &     3.8055E+02 &      3.240E-01 &      1.746E-01 \\
     9.5794E-01 &     1.1349E+03 &     3.8665E+02 &      3.251E-01 &      1.752E-01 \\
     9.6364E-01 &     1.1440E+03 &     3.9282E+02 &      3.262E-01 &      1.757E-01 \\
     9.6935E-01 &     1.1531E+03 &     3.9905E+02 &      3.273E-01 &      1.762E-01 \\
     9.7505E-01 &     1.1622E+03 &     4.0534E+02 &      3.284E-01 &      1.767E-01 \\
     9.8075E-01 &     1.1714E+03 &     4.1170E+02 &      3.295E-01 &      1.772E-01 \\
     9.8645E-01 &     1.1806E+03 &     4.1812E+02 &      3.306E-01 &      1.777E-01 \\
     9.9215E-01 &     1.1899E+03 &     4.2460E+02 &      3.316E-01 &      1.782E-01 \\
     9.9785E-01 &     1.1992E+03 &     4.3114E+02 &      3.327E-01 &      1.787E-01 \\
     1.0036E+00 &     1.2085E+03 &     4.3775E+02 &      3.337E-01 &      1.792E-01 \\
     1.0093E+00 &     1.2179E+03 &     4.4443E+02 &      3.348E-01 &      1.797E-01 \\
     1.0150E+00 &     1.2273E+03 &     4.5116E+02 &      3.358E-01 &      1.802E-01 \\
     1.0207E+00 &     1.2368E+03 &     4.5796E+02 &      3.368E-01 &      1.806E-01 \\
     1.0264E+00 &     1.2462E+03 &     4.6483E+02 &      3.378E-01 &      1.811E-01 \\
     1.0321E+00 &     1.2558E+03 &     4.7176E+02 &      3.388E-01 &      1.816E-01 \\
     1.0378E+00 &     1.2653E+03 &     4.7876E+02 &      3.398E-01 &      1.820E-01 \\
     1.0435E+00 &     1.2749E+03 &     4.8582E+02 &      3.408E-01 &      1.825E-01 \\
     1.0492E+00 &     1.2846E+03 &     4.9295E+02 &      3.418E-01 &      1.829E-01 \\
     1.0549E+00 &     1.2942E+03 &     5.0015E+02 &      3.427E-01 &      1.834E-01 \\
     1.0606E+00 &     1.3040E+03 &     5.0741E+02 &      3.437E-01 &      1.838E-01 \\
     1.0663E+00 &     1.3137E+03 &     5.1474E+02 &      3.446E-01 &      1.843E-01 \\
     1.0720E+00 &     1.3235E+03 &     5.2214E+02 &      3.456E-01 &      1.847E-01 \\
     1.0777E+00 &     1.3333E+03 &     5.2960E+02 &      3.465E-01 &      1.852E-01 \\
     1.0834E+00 &     1.3432E+03 &     5.3713E+02 &      3.474E-01 &      1.856E-01 \\
     1.0891E+00 &     1.3531E+03 &     5.4473E+02 &      3.484E-01 &      1.860E-01 \\
     1.0948E+00 &     1.3631E+03 &     5.5240E+02 &      3.493E-01 &      1.864E-01 \\
     1.1005E+00 &     1.3731E+03 &     5.6013E+02 &      3.502E-01 &      1.869E-01 \\
     1.1062E+00 &     1.3831E+03 &     5.6794E+02 &      3.511E-01 &      1.873E-01 \\
     1.1119E+00 &     1.3932E+03 &     5.7581E+02 &      3.519E-01 &      1.877E-01 \\
     1.1176E+00 &     1.4033E+03 &     5.8375E+02 &      3.528E-01 &      1.881E-01 \\
     1.1233E+00 &     1.4135E+03 &     5.9176E+02 &      3.537E-01 &      1.885E-01 \\
     1.1290E+00 &     1.4237E+03 &     5.9985E+02 &      3.546E-01 &      1.889E-01 \\
     1.1347E+00 &     1.4339E+03 &     6.0800E+02 &      3.554E-01 &      1.893E-01 \\
     1.1404E+00 &     1.4442E+03 &     6.1622E+02 &      3.563E-01 &      1.897E-01 \\
     1.1461E+00 &     1.4545E+03 &     6.2451E+02 &      3.571E-01 &      1.901E-01 \\
     1.1518E+00 &     1.4649E+03 &     6.3288E+02 &      3.580E-01 &      1.905E-01 \\
     1.1575E+00 &     1.4753E+03 &     6.4131E+02 &      3.588E-01 &      1.909E-01 \\
     1.1632E+00 &     1.4857E+03 &     6.4982E+02 &      3.596E-01 &      1.913E-01 \\
     1.1689E+00 &     1.4962E+03 &     6.5840E+02 &      3.604E-01 &      1.916E-01 \\
     1.1746E+00 &     1.5067E+03 &     6.6705E+02 &      3.612E-01 &      1.920E-01 \\
     1.1803E+00 &     1.5173E+03 &     6.7577E+02 &      3.620E-01 &      1.924E-01 \\
     1.1860E+00 &     1.5279E+03 &     6.8457E+02 &      3.628E-01 &      1.928E-01 \\
     1.1917E+00 &     1.5386E+03 &     6.9343E+02 &      3.636E-01 &      1.931E-01 \\
     1.1974E+00 &     1.5493E+03 &     7.0237E+02 &      3.644E-01 &      1.935E-01 \\
     1.2031E+00 &     1.5600E+03 &     7.1139E+02 &      3.652E-01 &      1.939E-01 \\
     1.2088E+00 &     1.5708E+03 &     7.2047E+02 &      3.659E-01 &      1.942E-01 \\
     1.2145E+00 &     1.5816E+03 &     7.2964E+02 &      3.667E-01 &      1.946E-01 \\
     1.2202E+00 &     1.5925E+03 &     7.3887E+02 &      3.675E-01 &      1.949E-01 \\
     1.2259E+00 &     1.6034E+03 &     7.4818E+02 &      3.682E-01 &      1.953E-01 \\
     1.2316E+00 &     1.6144E+03 &     7.5756E+02 &      3.690E-01 &      1.956E-01 \\
     1.2373E+00 &     1.6254E+03 &     7.6702E+02 &      3.697E-01 &      1.959E-01 \\
     1.2430E+00 &     1.6364E+03 &     7.7656E+02 &      3.704E-01 &      1.963E-01 \\
     1.2487E+00 &     1.6475E+03 &     7.8617E+02 &      3.712E-01 &      1.966E-01 \\
     1.2544E+00 &     1.6586E+03 &     7.9585E+02 &      3.719E-01 &      1.970E-01 \\
     1.2601E+00 &     1.6698E+03 &     8.0561E+02 &      3.726E-01 &      1.973E-01 \\
     1.2658E+00 &     1.6810E+03 &     8.1545E+02 &      3.733E-01 &      1.976E-01 \\
     1.2715E+00 &     1.6923E+03 &     8.2536E+02 &      3.740E-01 &      1.979E-01 \\
     1.2772E+00 &     1.7036E+03 &     8.3535E+02 &      3.747E-01 &      1.983E-01 \\
     1.2829E+00 &     1.7150E+03 &     8.4542E+02 &      3.754E-01 &      1.986E-01 \\
     1.2886E+00 &     1.7264E+03 &     8.5556E+02 &      3.761E-01 &      1.989E-01 \\
     1.2943E+00 &     1.7378E+03 &     8.6578E+02 &      3.768E-01 &      1.992E-01 \\
     1.3000E+00 &     1.7493E+03 &     8.7608E+02 &      3.774E-01 &      1.995E-01 \\
\hline
\end{longtable} 
\tablefoot{Table entries as in the previous table.} 
\label{tab_A2}

\end{document}